\documentclass[preprint,12pt]{elsarticle}

\usepackage{wrapfig}

\usepackage{graphics}
\usepackage{graphicx}
\usepackage{epsfig}
\usepackage{amssymb}
\usepackage{amsthm}
\usepackage{amsmath}

\usepackage[a4paper,left=2.0cm,right=2.0cm,top=2.3cm,bottom=2.3cm]{geometry}

\usepackage{caption}
\usepackage[labelfont=bf]{caption}
\usepackage[labelsep=space]{caption}
\usepackage[nodots, compress]{numcompress}
\biboptions{sort&compress}

\usepackage{xcolor}
\usepackage{hyperref}
\hypersetup{
colorlinks,
citecolor=[violet],
linkcolor=[red],
urlcolor=[blue]
}

\journal{Journal of Power Sources}

\usepackage{graphicx}
\usepackage{amsmath}
\begin{document}

\begin{frontmatter}

\title{Synergistic interface stability and high room-temperature ionic conductivity for wide-temperature all-solid-state batteries based on Li$_{6+x}$Si$_{x}$Sb$_{1-x}$S$_5$I electrolytes}

\author[1]{Liang Ming}
\author[1]{Qizhiran Sun}
\author[1]{Guanping Xu}
\author[1]{Muqing Su}
\author[1]{Enyan Zhao}
\author[1]{Wenzhe Gu}
\author[1]{Weng-Fu Io}
\author[1]{Kwun Nam Hui}

\author[2]{Chuang Yu\corref{cor1}}
\ead{cyu2020@hust.edu.cn}

\author[1]{Hai-Feng Li\corref{cor1}}
\ead{haifengli@um.edu.mo}

\cortext[cor1]{Corresponding authors.}
\affiliation[1]{organization={Institute of Applied Physics and Materials Engineering, University of Macau},
            addressline={Avenida da Universidade, Taipa}, 
            city={Macao S.A.R.},
            postcode={999078}, 
            country={China}}
\affiliation[2]{organization={School of Information Mechanics and Sensing Engineering, Xidian University},
            city={Xian PR},
            postcode={710126}, 
            country={China}}
      
\begin{abstract}
Solid-state lithium-ion batteries (LIBs) are increasingly recognized for their exceptional energy density and safety, yet their widespread adoption is challenged by persistent issues such as thermal and electrochemical instability, dendrite formation, and limited compatibility with high-voltage cathodes. Sulfide-based solid electrolytes (SEs), particularly iodide argyrodites, offer outstanding ionic conductivity and stability; however, practical application is constrained by the formation of space-charge layers, slow ion transport, and susceptibility to dendrite penetration. To address these challenges, we synthesize a novel Li$_{6.6}$Si$_{0.6}$Sb$_{0.4}$S$_5$I argyrodite electrolyte via ball milling and heat treatment, achieving a remarkable room-temperature ionic conductivity of 9.9~mS~cm$^{-1}$. The electrolyte is integrated with a LiNbO$_3$-coated LiNi$_{0.7}$Co$_{0.1}$Mn$_{0.2}$O$_2$ cathode to form an all-solid-state battery, which demonstrates an initial discharge capacity of 171.2~mAh~g$^{-1}$, retains 84.2\% of its capacity after 200 cycles at 0.5C, and maintains stable cycling across a broad temperature range (-20 $^\circ$C to 60 $^\circ$C). Our study shows that tailored electrolyte composition and a composite cathode configuration significantly enhance cycling stability and improve interfacial protection. These findings highlight the potential of Si-doped antimony-type iodide argyrodites for next-generation high-performance all-solid-state batteries, offering durable operation under diverse thermal conditions.
\end{abstract}


\begin{keyword}
Solid-state lithium batteries \sep Si-doped argyrodite electrolytes \sep Interface engineering \sep Ionic conductivity \sep Cycling stability
\end{keyword}

\end{frontmatter}

\section{Introduction}

Lithium-ion batteries (LIBs) are among the most promising energy-storage technologies, offering high energy density (3860~mAh~g$^{-1}$) and excellent charge retention. LIBs have achieved widespread adoption across modern applications and are integral to ongoing research, particularly with the rapid growth of electric vehicles~\cite{1,2,3,4,5,6,7}. Accordingly, considerable efforts have focused on improving cycle life, energy efficiency, safety, and battery design. However, safety concerns arising from the thermal and electrochemical instability of organic solvents, as well as intrinsic energy limitations, continue to hinder progress in liquid lithium-metal batteries. Solid electrolytes (SEs) have emerged as viable alternatives due to their nonflammability, chemical stability, and compatibility with lithium metal, which can suppress lithium dendrite growth~\cite{6,7,8,9,10, 11}. Among SEs, sulfide-based electrolytes are notable for their high room-temperature (RT) ionic conductivity (10$^{-4}$--10$^{-2}$~S~cm$^{-1}$), soft mechanical properties, and straightforward synthesis~\cite{12,13,14}. Nevertheless, broader implementation is constrained by three challenges: (i) formation of a space-charge layer at the cathode--electrolyte interface; (ii) sluggish ion transport due to thick or poorly optimized electrolyte layers; and (iii) lithium dendrite penetration along internal grain boundaries~\cite{15,16,17}.

To address these limitations, research has focused on Li-argyrodite $\mathrm{Li}_6\mathrm{PS}_5\mathrm{X}$ (X = Cl, Br, I), which has attracted significant attention for its satisfactory ionic conductivity and electrochemical stability~\cite{18,19,20,21,22}. Among these, $\mathrm{Li}_6\mathrm{PS}_5\mathrm{I}$ exhibits relatively low ionic conductivity because the large radius of $\mathrm{I}^-$ within the $\mathrm{PS}_4^{3-}$ framework promotes a more ordered arrangement and reduces $\mathrm{I}^-/\mathrm{S}^{2-}$ site disorder~\cite{23,24,25}. Even so, the larger lattice volume and structural robustness of $\mathrm{Li}_6\mathrm{PS}_5\mathrm{I}$ enable optimization via elemental substitution without substantially altering the lattice. Enhancing the ionic conductivity and stability of $\mathrm{Li}_6\mathrm{PS}_5\mathrm{I}$ typically requires improved synthesis strategies. Aliovalent substitution at the P site (Si, Sn, Ge) and chalcogenide substitution at the S site have been explored~\cite{26,27,28}. Post-synthesis heat treatment can increase ionic conductivity to 3.6$\times$10$^{-3}$~S~cm$^{-1}$, indicating that Si doping establishes favorable $\mathrm{Li}^+$ transport pathways~\cite{29,30}. Nonetheless, structural constraints in $\mathrm{Li}_6\mathrm{PS}_5\mathrm{I}$ limit further optimization. Consequently, research has shifted toward antimony-type $\mathrm{Li}_6\mathrm{SbS}_5\mathrm{I}$, which offers a larger lattice volume and more $\mathrm{Li}^+$ sites. Compared with $\mathrm{Li}_6\mathrm{PS}_5\mathrm{I}$, the expanded lattice of $\mathrm{Li}_6\mathrm{SbS}_5\mathrm{I}$ provides additional defect sites and facilitates more $\mathrm{Li}^+$ transport pathways, thereby reducing the migration energy barrier without significant lattice distortion~\cite{31,32,33,34,35}. A new family of iodide argyrodites, such as $\mathrm{Li}_{6.6}\mathrm{Si}_{0.6}\mathrm{Sb}_{0.4}\mathrm{S}_5\mathrm{I}$ reported by Zhou~\cite{36}, demonstrates the promise of Si-doped electrolytes due to their low cost and favorable effects on performance~\cite{37}. However, few studies have addressed their practical application in all-solid-state batteries, mainly because of poor electrochemical stability with high-voltage commercial cathodes~\cite{38,39,40,41}.

Despite progress in developing high-conductivity sulfide electrolytes, their integration with high-voltage cathodes remains a central challenge for practical all-solid-state batteries. Commercial layered oxide cathodes, such as LiNi$_x$Co$_y$Mn$_z$O$_2$, offer high capacity and robust cyclability in conventional liquid-electrolyte systems~\cite{42,43,44,45,46}, but their direct contact with sulfide electrolytes can induce space-charge-layer formation, transition-metal interdiffusion, oxidative electrolyte decomposition, and increased interfacial resistance~\cite{47,48,49,50,51}. To mitigate these issues, a common strategy is to construct a physical barrier between the cathode and SE, thereby preventing inter-diffusion and the formation of resistive interfaces~\cite{52,53,54,55,56,57,58}. Surface coatings, including $\mathrm{LiNbO}_3$~\cite{59} and $\mathrm{Li}_2\mathrm{ZrO}_3$~\cite{60,61} have therefore been widely used to construct a physical and electrochemical barrier between oxide cathodes and sulfide electrolytes. However, coating layers alone cannot fully resolve the transport requirements of composite cathodes, where ionic conduction through the solid electrolyte and electronic conduction through conductive additives must be simultaneously regulated. In particular, carbon additives are necessary for establishing an electronic percolation network, but excessive carbon may create continuous electron-leakage pathways and accelerate oxidative decomposition of sulfide electrolytes. Therefore, the key issue is not simply to improve electrolyte conductivity or introduce cathode coatings independently, but to coordinate bulk ion transport, interfacial protection, and electronic-conduction regulation within a unified composite-cathode design.

In this study, we establish an electrolyte--interface co-design strategy for wide-temperature sulfide-based all-solid-state batteries. The principal novelty of this work lies in integrating three closely coupled design elements into one coherent framework: Si substitution in Sb-based iodide argyrodite is used to enhance bulk Li-ion transport; LiNbO$_3$ coating is employed to mitigate direct interfacial reactions between the high-voltage LiNi$_{0.7}$Co$_{0.1}$Mn$_{0.2}$O$_2$ cathode and the sulfide electrolyte; and the VGCF content is optimized to balance electronic percolation with interfacial stability. Through ball milling followed by heat treatment, the optimized Li$_{6.6}$Si$_{0.6}$Sb$_{0.4}$S$_5$I electrolyte achieves a high room-temperature ionic conductivity of $9.9 \times 10^{-3}$ S cm$^{-1}$ and a low activation energy of 0.18 eV. When paired with a LiNbO$_3$-coated LiNi$_{0.7}$Co$_{0.1}$Mn$_{0.2}$O$_2$ composite cathode containing an optimized conductive-carbon content, the assembled all-solid-state battery delivers an initial discharge capacity of 174.8 mAh g$^{-1}$ and retains 68.2\% of its capacity after 300 cycles at 0.5C. The cell further maintains stable cycling from $-20~^\circ$C to $60~^\circ$C. Structural characterization, electrochemical analysis, and first-principles calculations collectively reveal that the improved performance arises from the synergistic regulation of electrolyte composition and cathode interfacial architecture. This work therefore advances the current state of the art by demonstrating that high ionic conductivity must be coupled with controlled interfacial electronic transport to achieve stable, wide-temperature sulfide-based all-solid-state batteries.

\section{Results and Discussion}

\subsection{Structural and morphological evolution of Li$_{6+x}$Si$_{x}$Sb$_{1-x}$S$_5$I electrolytes}

Understanding the structural and morphological characteristics of Li$_{6+x}$Si$_{x}$Sb$_{1-x}$S$_5$I ($x = 0$, 0.6, 0.75) electrolytes is crucial for optimizing their performance in energy storage applications. High-energy ball milling was employed to synthesize these materials. To elucidate structural variations resulting from Si$^{4+}$ substitution, XRD patterns (Fig.~\ref{Fig. S1}) demonstrate that the synthesized samples possess highly argyrodite-type structures with $F$-$43m$ space group \cite{61,62}. In this process, the lightweight, low-cost, and non-toxic silicon element replaces a significant portion of Sb at the crystalline 16e Wykoff site, enhancing material sustainability and potential scalability. As shown in Fig.~\ref{Fig. S1}, the characteristic peaks near 30$^\circ$ and 50$^\circ$ gradually shift to higher angles with increasing Si substitution, which is attributed to the smaller ionic radius of Si compared to Sb. Minor Li$_2$S, LiI, and Li$_4$SiS$_4$ impurities were detected in Si-doped electrolytes upon reaching the solubility limit at $x = 0.6$. The quantity of impurities increases in the XRD patterns of Li$_{6.75}$Si$_{0.75}$Sb$_{0.25}$S$_5$I, indicating the accumulation of lithium-containing phases. The diffraction peaks of Li$_{6.75}$Si$_{0.75}$Sb$_{0.25}$S$_5$I shift to larger angles with higher Si substitution, suggesting lattice compression \cite{63}. These findings highlight the impact of Si doping on crystal structure and phase composition, providing a foundation for further analysis.

\begin{figure*}
\centering
\includegraphics[width=0.92\textwidth]{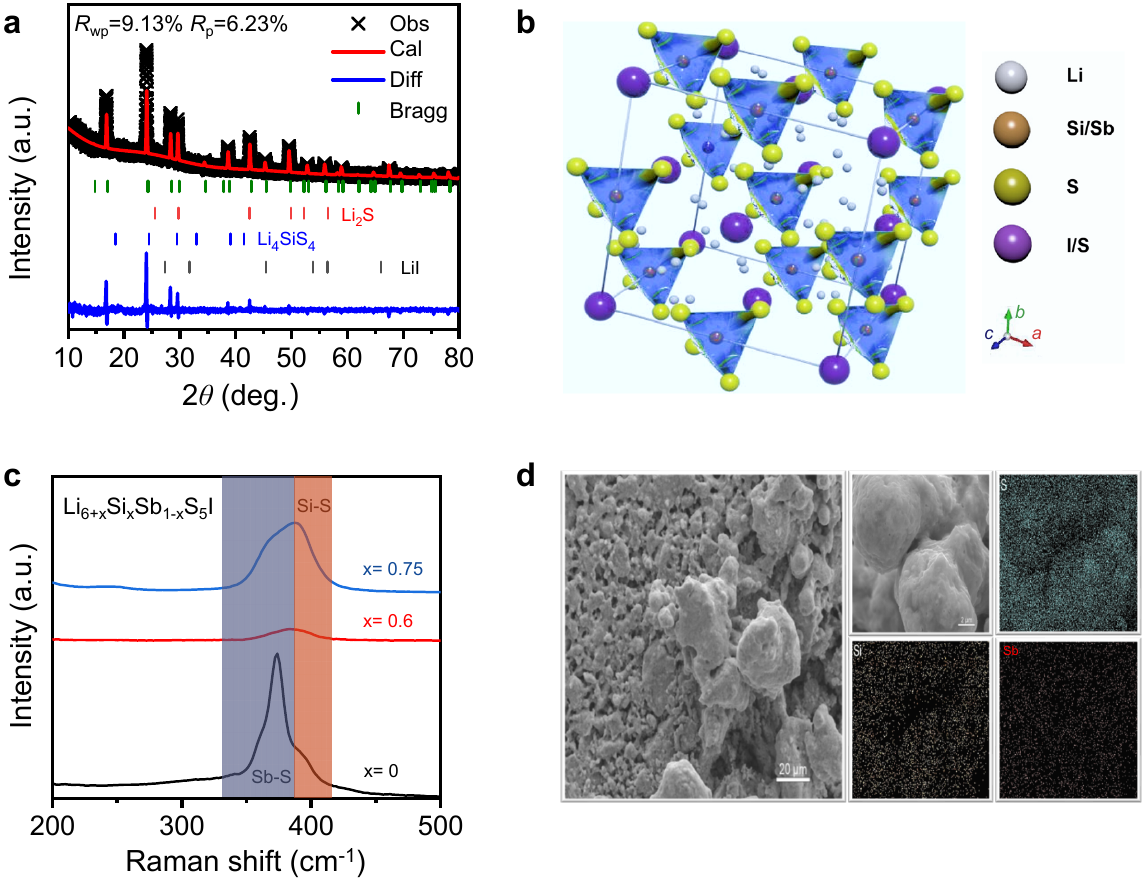}
\caption{
Structural characterization of Li$_{6+x}$Si$_{x}$Sb$_{1-x}$S$_5$I electrolytes. 
(a) XRD Rietveld refinement profiles, (b) illustration of the crystal structure, and (c) Raman spectra of Li$_{6+x}$Si$_{x}$Sb$_{1-x}$S$_5$I. (d) SEM image and corresponding EDS elemental mapping for Si, Sb, and S in Li$_{6.6}$Si$_{0.6}$Sb$_{0.4}$S$_5$I electrolyte, demonstrating uniform distribution of each element.
}
\label{Fig. 1}
\end{figure*}

Additionally, X-ray Rietveld refinement was performed using GSAS software to investigate the impact of aliovalent substitution in Li$_6$SbS$_5$I. The lattice parameters, atomic positions, occupation factors, and quantitative phase fractions for Li$_{6+x}$Si$_x$Sb$_{1-x}$S$_5$I ($x = 0$, 0.6, and 0.75) are summarized in Tables~\ref{tab:S1}--\ref{tab:S4}, respectively. The refinement results confirm the successful incorporation of Si$^{4+}$ into Li$_6$SbS$_5$I, as evidenced by the reduced lattice volume and lattice constants ($a$, $b$, and $c$) after annealing at 450~$^\circ$C with increasing Si content (Fig.~\ref{Fig. 1} and Fig.~\ref{Fig. S2}). The quantitative phase analysis further reveals the influence of Si substitution on phase purity. The optimized Li$_{6.6}$Si$_{0.6}$Sb$_{0.4}$S$_5$I sample contains 95.0~wt\% main phase with only minor impurity phases, including 2.0~wt\% Li$_2$S, 2.0~wt\% LiI, and 1.0~wt\% Li$_4$SiS$_4$. In contrast, for Li$_{6.75}$Si$_{0.75}$Sb$_{0.25}$S$_5$I, the main phase fraction decreases to 88.0~wt\%, while the fractions of Li$_2$S, LiI, and Li$_4$SiS$_4$ increase to 3.9~wt\%, 4.7~wt\%, and 3.4~wt\%, respectively. The increased impurity content at higher Si substitution levels may disrupt the continuous Li$^+$ transport network, increase grain-boundary resistance, and contribute to the observed decrease in ionic conductivity. Therefore, the Rietveld refinement results provide a quantitative explanation for the optimized conductivity observed in Li$_{6.6}$Si$_{0.6}$Sb$_{0.4}$S$_5$I.

Raman spectroscopy was conducted on the electrolytes to further verify Si substitution in Li$_{6+x}$Si$_{x}$Sb$_{1-x}$S$_5$I. As shown in Fig.~\ref{Fig. 1}(c), two distinct peaks at 362.6 and 384.8~cm$^{-1}$ are associated with Sb-S bonds in Li$_6$SbS$_5$I. In contrast, the peak around 362.6~cm$^{-1}$ broadens with increasing Si content, while a newly emerged peak at 394.0~cm$^{-1}$, attributed to Si-S bonds, provides additional evidence of partial Si doping for Sb \cite{64}. These Raman spectra are consistent with the XRD analysis. This complementary spectroscopic evidence reinforces the structural findings obtained from XRD and Rietveld refinement, confirming the successful incorporation of Si into the lattice.

SEM and EDS were applied to Li$_{6.6}$Si$_{0.6}$Sb$_{0.4}$S$_5$I. As depicted in Fig.~\ref{Fig. 1}(d), the Li$_{6.6}$Si$_{0.6}$Sb$_{0.4}$S$_5$I electrolyte exhibits an average particle size of $\sim$20~$\mu$m with homogeneous distribution of Si, Sb, and S. Such uniform morphology and elemental dispersion are advantageous for ensuring consistent electrochemical performance in practical applications. This observation provides a meaningful link between structural integrity and functional outcomes, paving the way for future studies on electrochemical behavior.

\subsection{Electrochemical properties and stability of Li$_{6+x}$Si$_{x}$Sb$_{1-x}$S$_5$I electrolytes}

\begin{figure*}
\centering
\includegraphics[width=0.92\textwidth]{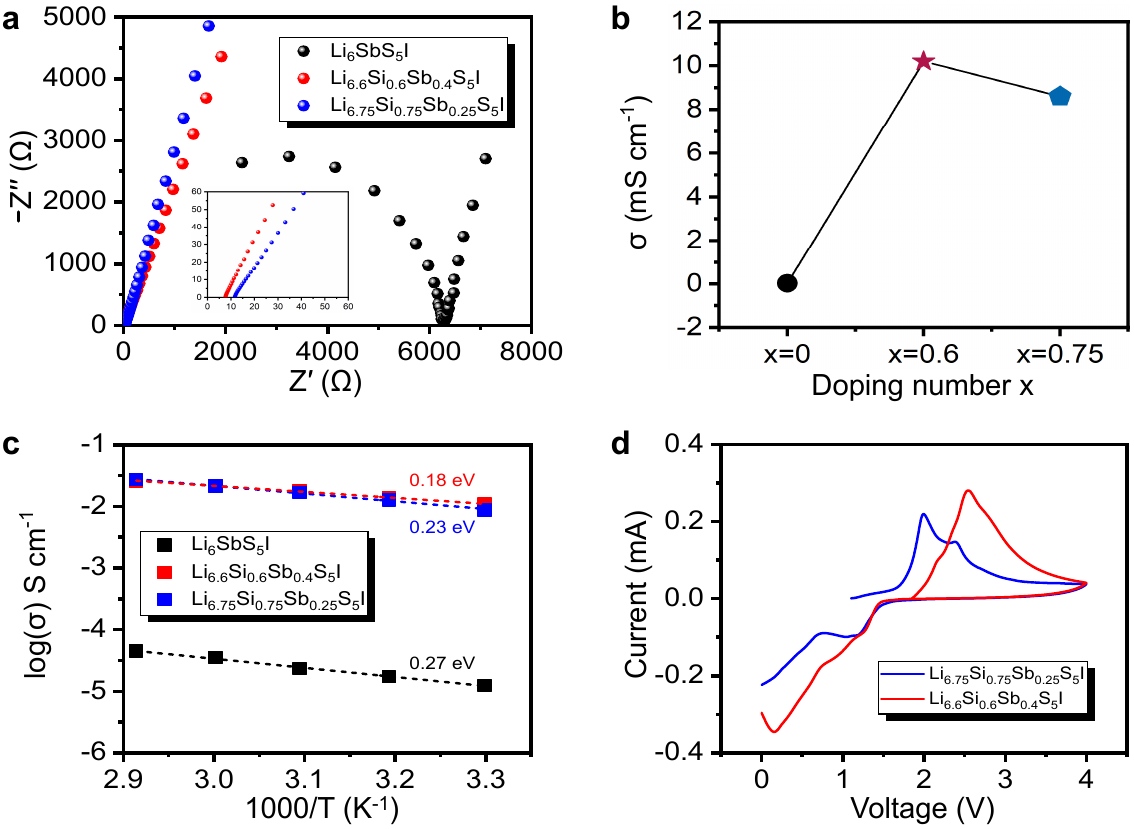}
\caption{Ionic conductivity and electrochemical stability analysis. 
(a) Nyquist plots, and (b) room-temperature ionic conductivity of Li$_6$SbS$_5$I, Li$_{6.6}$Si$_{0.6}$Sb$_{0.4}$S$_5$I, and Li$_{6.75}$Si$_{0.75}$Sb$_{0.25}$S$_5$I. (c) Arrhenius plots for the three electrolytes, revealing activation energies. (d) CV profiles of Li$_{6.6}$Si$_{0.6}$Sb$_{0.4}$S$_5$I and Li$_{6.75}$Si$_{0.75}$Sb$_{0.25}$S$_5$I, highlighting their electrochemical stability windows.
}
\label{Fig. 2}
\end{figure*}

Electrochemical characterization is essential for understanding the performance of Li$_{6+x}$Si$_{x}$Sb$_{1-x}$S$_5$I electrolytes in advanced energy storage systems. The Li-ion conductivities of cold-pressed Li$_{6+x}$Si$_{x}$Sb$_{1-x}$S$_5$I pellets at room temperature were systematically investigated using EIS in symmetric cells configured as blocking steel/Li$_{6+x}$Si$_{x}$Sb$_{1-x}$S$_5$I/blocking steel, as illustrated in Fig.~\ref{Fig. 2}(a) and Fig.~\ref{Fig. 2}(b). The as-prepared Li$_{6}$SbS$_5$I exhibits an ionic conductivity of $1.5 \times 10^{-5}$ S cm$^{-1}$ at 25~$^\circ$C, which is comparable to that of Li$_{6}$PS$_5$I ($1.36 \times 10^{-5}$ S cm$^{-1}$)~\cite{36}. Substituting Sb with Si significantly increases the ionic conductivity, with Li$_{6.6}$Si$_{0.6}$Sb$_{0.4}$S$_5$I reaching $9.9 \times 10^{-3}$ S cm$^{-1}$ at 25~$^\circ$C. However, when the Si substitution level is further increased to 0.75, exceeding the optimized substitution level, the ionic conductivity decreases to $8.1 \times 10^{-3}$ S cm$^{-1}$, as shown in Fig.~\ref{Fig. 2}(b).

To provide more direct experimental evidence for Li$^+$ transport kinetics, temperature-dependent EIS measurements were further performed for Li$_{6}$SbS$_5$I, Li$_{6.6}$Si$_{0.6}$Sb$_{0.4}$S$_5$I, and Li$_{6.75}$Si$_{0.75}$Sb$_{0.25}$S$_5$I over the temperature range of 30--70~$^\circ$C, as shown in Fig.~\ref{Fig. S3}(a-c). The impedance spectra were fitted using the same equivalent-circuit model of $R_s$-($R_{\mathrm{SE}} \parallel \mathrm{CPE}_{\mathrm{SE}}$)-$\mathrm{CPE}_{\mathrm{SS}}$, where $R_{\mathrm{SE}}$ was used as the effective ion-transport resistance of the solid electrolyte. Although the impedance responses of different samples show different features, especially the more distinguishable depressed semicircle in Fig.~\ref{Fig. S3}(c), all measurements were conducted using the same SS $\vert$ electrolyte $\vert$ SS blocking-cell configuration (Fig.~\ref{Fig. S3}(d)). Therefore, the fundamental electrochemical processes are consistent, and the spectral differences are mainly reflected by the fitted $R_{\mathrm{SE}}$ and CPE parameters. Since the bulk and grain-boundary contributions cannot be unambiguously separated for all samples, $R_{\mathrm{SE}}$ was used for the ionic conductivity and activation energy calculations.

The impedance of all samples decreases with increasing temperature, confirming the thermally activated nature of Li$^+$ conduction in these sulfide electrolytes. The ionic conductivity was calculated according to $\sigma = L/(R \times A)$, where $L$ is the pellet thickness, $R$ is the fitted $R_{\mathrm{SE}}$, and $A$ is the effective electrode area. Based on the temperature-dependent conductivity values extracted from Fig.~\ref{Fig. S3}(e), the activation energies were calculated using the Arrhenius equation, $\ln(\sigma T) = \ln A_0 - E_a/(k_B T)$, and the corresponding fitting results are shown in Fig.~\ref{Fig. 2}(c). The activation energies are 0.28 eV for Li$_{6}$SbS$_5$I, 0.18 eV for Li$_{6.6}$Si$_{0.6}$Sb$_{0.4}$S$_5$I, and 0.23 eV for Li$_{6.75}$Si$_{0.75}$Sb$_{0.25}$S$_5$I. Among these compositions, Li$_{6.6}$Si$_{0.6}$Sb$_{0.4}$S$_5$I demonstrates the highest ionic conductivity and the lowest activation energy, indicating the most favorable Li$^+$ transport kinetics. All the fitted parameters and data were listed in Tables~\ref{tab:S5}--\ref{tab:S7}

The reduced activation energy of Li$_{6.6}$Si$_{0.6}$Sb$_{0.4}$S$_5$I experimentally confirms that moderate Si substitution facilitates Li$^+$ migration in the argyrodite framework. The aliovalent substitution of Si enhances local structural disorder within the I$^-$/S$^{2-}$ framework and increases the concentration of mobile Li$^+$ ions, allowing additional Li ions to occupy energetically favorable migration-related sites. These structural modifications promote continuous Li$^+$ transport pathways and are consistent with the DFT-calculated reduction in Li$^+$ migration barriers, thereby supporting the conductivity enhancement mechanism.

To further evaluate the ion-transport behavior of the optimized composition, the lithium-ion transference number ($t_{\mathrm{Li}^+}$) of Li$_{6.6}$Si$_{0.6}$Sb$_{0.4}$S$_5$I was measured by DC polarization chronoamperometry (CA) in combination with EIS using a Li|Li$_{6.6}$Si$_{0.6}$Sb$_{0.4}$S$_5$I|Li symmetric cell. As shown in Fig.~\ref{Fig. S4}(a,b), the impedance spectra recorded before and after polarization, together with the current response under a constant DC bias, yield a $t*{\mathrm{Li}^+}$ value of 0.91. This relatively high transference number indicates that Li$^+$ dominates the ionic transport in the electrolyte, which is beneficial for alleviating concentration polarization and promoting fast Li$^+$ migration. Combined with its highest ionic conductivity and lowest activation energy, Li$_{6.6}$Si$_{0.6}$Sb$_{0.4}$S$_5$I exhibits superior ion-transport characteristics among the investigated compositions.

Current-voltage (CV) measurements were conducted to evaluate the electrochemical stability of Li$_{6+x}$Si$_{x}$Sb$_{1-x}$S$_5$I within a potential range of 0.0–4.0~V at a scan rate of 0.1~mV~s$^{-1}$, as depicted in Fig.~\ref{Fig. 2}(d). The battery configuration comprised Li$_{6+x}$Si$_{x}$Sb$_{1-x}$S$_5$I + VGCF/Li$_{6+x}$Si$_{x}$Sb$_{1-x}$S$_5$I/Li-In. No redox peaks were detected for Li$_6$SbS$_5$I, likely due to its low ionic conductivity. Notably, Li$_{6.6}$Si$_{0.6}$Sb$_{0.4}$S$_5$I exhibits an electrochemical stability window up to 2.65~V (vs. Li–In), which is considerably broader than that of Li$_{6.75}$Si$_{0.75}$Sb$_{0.25}$S$_5$I (1.98~V vs. Li–In). The combined advantages of high ionic conductivity and a wide electrochemical stability window highlight the promise of Li$_{6.6}$Si$_{0.6}$Sb$_{0.4}$S$_5$I for advanced energy storage applications.

\begin{figure*}
\centering
\includegraphics[width=0.92\textwidth]{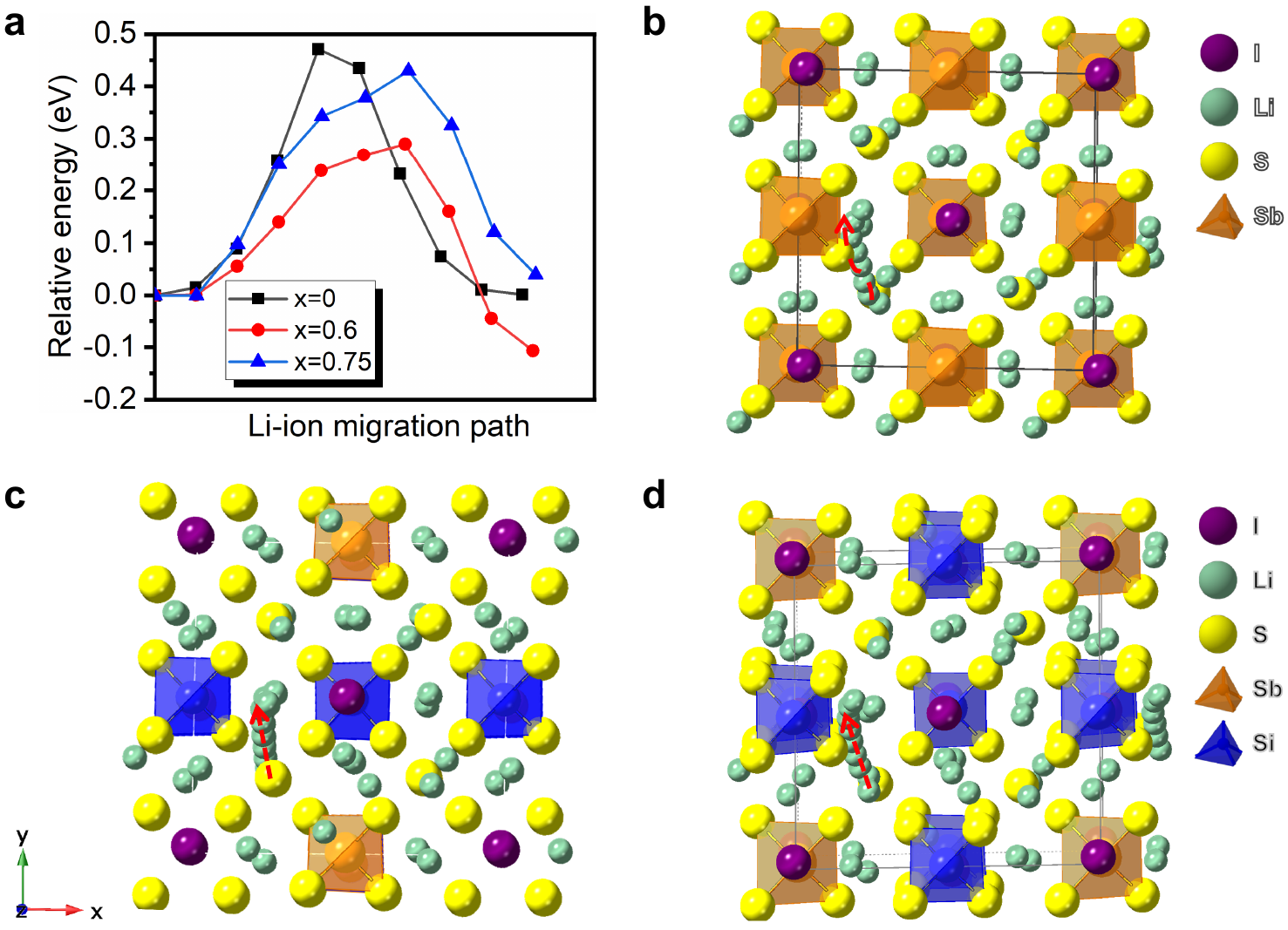}
\caption{Li-ion migration barriers and pathways for various electrolytes. 
(a) Li-ion migration barriers and the corresponding path diagrams for (b) Li$_6$SbS$_5$I, (c) Li$_{6.6}$Si$_{0.6}$Sb$_{0.4}$S$_5$I, and (d) Li$_{6.75}$Si$_{0.75}$Sb$_{0.25}$S$_5$I.
}
\label{Fig. 3}
\end{figure*}

To elucidate the impact of Si doping on Li$^+$ transport within the Li$_6$SbS$_5$I system, spin-polarized density functional theory (DFT) calculations were performed. After identifying the thermodynamically stable structures of Li$_6$SbS$_5$I, Li$_{6.6}$Si$_{0.6}$Sb$_{0.4}$S$_5$I, and Li$_{6.75}$Si$_{0.75}$Sb$_{0.25}$S$_5$I, representative Li-ion migration pathways were selected for each composition. The migration energy profiles were constructed by interpolating multiple intermediate images between the initial and final states, with the highest-energy transition state taken as the saddle point. Using the initial state energy as a reference, the migration barriers for different doping levels were directly compared. As shown in Fig.~\ref{Fig. 3}(a), the calculated barriers exhibit a clear doping dependence: Li$_6$SbS$_5$I shows a barrier of $\sim$0.47~eV, which decreases to $\sim$0.28~eV in Li$_{6.6}$Si$_{0.6}$Sb$_{0.4}$S$_5$I and then rises to $\sim$0.42~eV in Li$_{6.75}$Si$_{0.75}$Sb$_{0.25}$S$_5$I. The lowest barrier ($\sim$0.28~eV) observed for Li$_{6.6}$Si$_{0.6}$Sb$_{0.4}$S$_5$I indicates the most favorable Li-ion mobility among the three systems. This trend suggests that moderate Si doping significantly facilitates Li-ion diffusion, while excessive doping reintroduces migration hindrance. The corresponding migration pathways are visualized in Fig.~\ref{Fig. 3}(b) for Li$_6$SbS$_5$I, Fig.~\ref{Fig. 3}(c) for Li$_{6.6}$Si$_{0.6}$Sb$_{0.4}$S$_5$I, and Fig.~\ref{Fig. 3}(d) for Li$_{6.75}$Si$_{0.75}$Sb$_{0.25}$S$_5$I, illustrating how local structural changes induced by doping reshape the energy landscape for Li-ion transport.

\subsection{Optimized cathode composition for enhanced performance in SSBs}

\begin{figure*}
\centering
\includegraphics[width=0.92\textwidth]{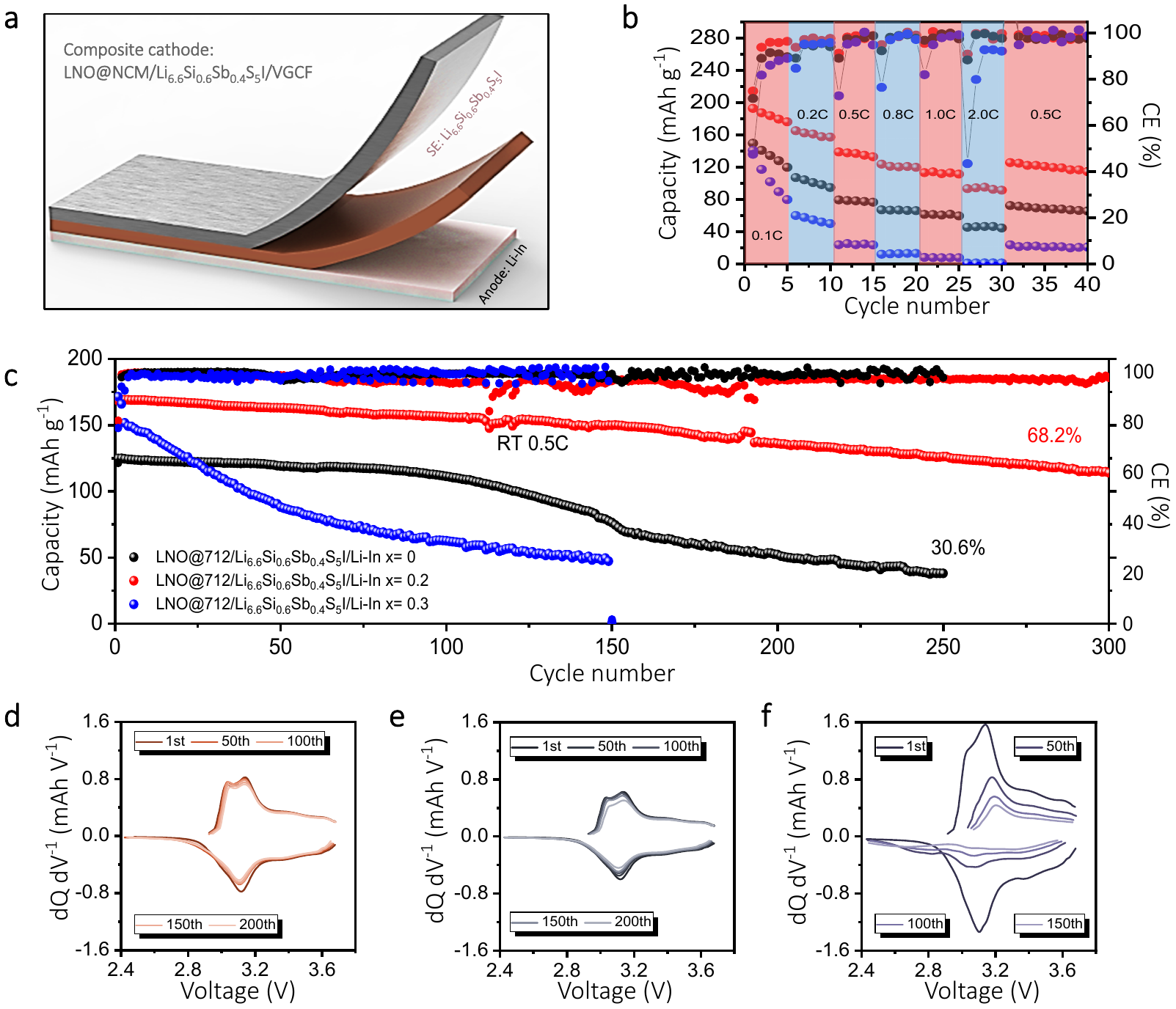}
\caption{Electrochemical performance of composite cathodes. 
(a) Schematic illustration of the assembled solid-state batteries. (b) Rate capability and (c) long-term cycling performance of batteries with different carbon ratios. (d--f) Corresponding d$Q$/d$V^{-1}$ plots for batteries with carbon ratios of (d) $x = 0.2$, (e) $x = 0$, and (f) $x = 0.3$, revealing the effect of carbon content on redox processes.
}
\label{Fig. 4}
\end{figure*}

To assess the practical applicability of Li$_{6.6}$Si$_{0.6}$Sb$_{0.4}$S$_5$I electrolytes, all-solid-state batteries (SSBs) were fabricated using LiNbO$_3$@LiNi$_{0.7}$Co$_{0.1}$Mn$_{0.2}$O$_2$ (LNO@NCM) as the cathode active material, Li$_{6.6}$Si$_{0.6}$Sb$_{0.4}$S$_5$I as the solid electrolyte, and Li--In as the anode, cycled between 2.4 and 3.7~V (vs. Li--In). Fig.~\ref{Fig. 4}(a) presents the cathode modification and battery configuration, emphasizing the integration strategy of high-conductivity Li$_{6.6}$Si$_{0.6}$Sb$_{0.4}$S$_5$I electrolytes with modified oxide cathodes to achieve robust electrochemical performance. To further verify the effectiveness of the cathode surface modification in stabilizing the cathode/electrolyte interface, the morphology and elemental distribution of the LNO coating layer were examined by SEM and EDS mapping. As shown in Fig.~\ref{Fig. S5}, the SEM and elemental mapping results clearly demonstrate that the LNO coating on the NCM cathode is uniform and conformal across the secondary particle surfaces. In the Ni, Co, and Mn maps, the primary NCM particles are homogeneously distributed, while the Nb map shows a continuous thin layer ($\sim$ 50--70~nm) enveloping the NCM aggregates, indicating that the LNO is well-adhered. This thin and uniform coating ensures that the electrolyte predominantly contacts the LNO layer rather than the NCM active material, thereby suppressing parasitic side reactions at the cathode/electrolyte interface and enhancing interfacial stability during electrochemical cycling.

Beyond the interfacial protection provided by the LNO coating, the conductive carbon content in the composite cathode must also be carefully optimized to achieve balanced ion/electron transport and interfacial stability. Although the LNO layer suppresses direct contact between NCM and the sulfide electrolyte, insufficient conductive carbon leads to incomplete electronic percolation and increased polarization, whereas excessive carbon may enlarge the electronically conductive contact area with the sulfide electrolyte and accelerate oxidative side reactions. Therefore, to clarify the role of conductive carbon content in the LNO-modified composite cathode, cathodes with different VGCF contents of 0 wt\%, 2 wt\%, and 3 wt\% were systematically investigated. This optimization is essential for coupling the chemically protective LNO coating with an efficient electronic conduction network, thereby enabling stable and high-rate operation of the composite cathode.

Fig.~\ref{Fig. 4}(b) displays the rate performance results. LNO@NCM ($x = 0$)/Li$_{6.6}$Si$_{0.6}$Sb$_{0.4}$S$_5$I/Li-In delivered discharge capacities of 157~mAh~g$^{-1}$ at 0.1C, 115.6~mAh~g$^{-1}$ at 0.2C, 80.1~mAh~g$^{-1}$ at 0.5C, 67.5~mAh~g$^{-1}$ at 0.8C, 64.7~mAh~g$^{-1}$ at 1.0C, and 43.2~mAh~g$^{-1}$ at 2.0C. In contrast, the LNO@NCM-VGCF ($x = 0.3$)/Li$*{6.6}$Si$*{0.6}$Sb$*{0.4}$S$*5$I/Li-In exhibited markedly low discharge capacities across all C-rates, approaching zero at 2.0C. The optimized cathode design, LNO@NCM-VGCF ($x = 0.2$)/Li$*{6.6}$Si$*{0.6}$Sb$_{0.4}$S$_5$I/Li-In, achieved substantially higher discharge capacities: 197.6, 173.5, 141.3, 130.7, 121.4, and 101.7~mAh~g$^{-1}$ at 0.1C, 0.2C, 0.5C, 0.8C, 1.0C, and 2.0C, respectively. These results highlight the significance of cathode composition in optimizing battery performance.

Beyond rate capability, cycling stability is crucial for practical applications. The enhanced rate capability of the optimized battery is attributed to improved electronic conductivity from suitable carbon additives in the cathode mixture and the mitigation of the space-charge-layer effect \cite{65}. Additionally, sulfide electrolytes participate in electrochemical reactions within the cathode mixture, contributing to capacity in the presence of carbon additives. As depicted in Fig.~\ref{Fig. 4}(c) and Fig.~\ref{Fig. S6}(c), SSBs with a higher carbon-additive content ($x=0.3$) delivered the highest discharge capacity of 179.3 mAh g$^{-1}$ at 0.5C, but exhibited rapid capacity decay, with the capacity retention dropping to 10.3\% after 100 cycles, indicating a pronounced space-charge-layer effect. With identical cathode loading of 4.46 mg cm$^{-2}$, SSBs utilizing LNO@NCM without carbon ($x=0$) showed lower discharge capacity and more pronounced polarization at 0.5C, with the capacity retention decreasing to 30.6\% after 250 cycles. In contrast, SSBs with a moderate carbon-additive content ($x=0.2$), as shown in Fig.~S6(b) and Fig.~\ref{Fig. 4}(c), exhibited an initial discharge capacity of 174.8 mAh g$^{-1}$ and a capacity retention of 68.2\% after 300 cycles, demonstrating the advantages of optimized carbon content for both rate performance and long-term cycling stability.

To further elucidate the underlying mechanisms, differential capacity (d$Q$/d$V$) analysis was performed, clarifying the impact of carbon content on cathode reversibility and phase stability (Figs.~\ref{Fig. 4}(d)-(f)). The carbon-free cathode ($x = 0$) exhibited moderately reversible redox peaks with minimal polarization, suggesting limited interfacial side reactions at the LNO@NCM/electrolyte interface \cite{66}. However, the absence of a conductive network restricted electron transport, resulting in insufficient capacity utilization. A gradual shift of the oxidation peak to higher voltages upon cycling indicated ongoing accumulation of side reactions, progressively reducing capacity \cite{67,68}. Conversely, the high-carbon cathode ($x = 0.3$) initially delivered high capacity but suffered severe capacity fading, accompanied by intensified H1/M and H2/H3 phase transition signals and significant voltage hysteresis. This behavior is attributed to excessive electronic conductivity, which accelerates interfacial decomposition and promotes irreversible phase transitions \cite{69}. Notably, the optimally formulated cathode ($x = 0.2$) achieved a balanced trade-off, maintaining strong redox peak intensities with negligible voltage shift over 200 cycles. The well-preserved H1/M and H2/H3 phase transitions, along with minimal peak decay, confirm suppressed interfacial degradation and stable lithium (de)intercalation kinetics, emphasizing the critical role of carbon modulation in enabling both high capacity and prolonged cyclability \cite{70}.

Comparatively, the minimal reduction in peak intensity and the consistent overlap of differential capacity plots for the $x = 0.2$ cathode suggest that the optimized mixture effectively suppresses voltage reduction and side reactions after 200 cycles \cite{71}. These results indicate that LiNbO$_3$ coating layers and carbon additives ($x = 0.2$) are effective in enhancing the reversibility of H1/M and H2/H3 phase transitions. According to Figs.~\ref{Fig. 4}(d)--(f), the enhanced reversibility observed in the H1/M (2.8--3.2~V) and H2/H3 (3.5--3.7~V) regions underscores the importance of optimized cathode configuration in mitigating polarization and promoting efficient Li-ion diffusion. These findings provide compelling evidence for the efficacy of NCM modifications, particularly in improving phase transition reversibility during cycling.

Impedance analysis of LNO@NCM712/Li$_{6.6}$Si$_{0.6}$Sb$_{0.4}$S$_5$I/Li-In cells after extended cycling at 0.5C reveals significant insights into interfacial evolution and charge transport behavior (see Fig.~\ref{Fig. S7}). As summarized in Table~\ref{tab:S8}, the optimized cathode formulation with $x = 0.2$ carbon additive demonstrates superior interfacial stability and lower overall impedance compared to other configurations. The bulk electrolyte resistance ($R_{\text{SE\_bulk}}$) for the $x = 0.2$ configuration is 15.5~$\Omega$, substantially lower than the carbon-free ($x = 0$, 39.4~$\Omega$) and high-carbon ($x = 0.3$, 30.1~$\Omega$) counterparts, suggesting that moderate carbon content enhances ionic percolation while preserving electrolyte integrity \cite{72}. The cathode/electrolyte interface resistance follows a similar trend, with $x = 0.2$ exhibiting the lowest value (8.8~$\Omega$), indicative of effective suppression of space charge layer formation and stable interfacial contact. Notably, the anode/electrolyte interface resistance for $x = 0.2$ (121~$\Omega$) is significantly reduced compared to $x = 0$ (357~$\Omega$) and $x = 0.3$ (247~$\Omega$), reflecting improved Li-In alloy stability and reduced interfacial degradation \cite{73}. Collectively, this comprehensive impedance analysis confirms that the $x = 0.2$ cathode configuration achieves an optimal balance between electronic conductivity and interfacial stability, thereby contributing to the superior cycling performance observed in this study.

\subsection{Mechanistic insights into interfacial evolution and stability in advanced solid-state batteries}

To gain deeper insights into the interfacial evolution during electrochemical cycling, we characterized the surface morphology of the cathodes before and after cycling using SEM. To ensure stable electrochemical reaction kinetics, the cells were tested at a low current rate of 0.2C. As shown in Fig.~\ref{Fig. S8}, the LNO@NCM-VGCF ($x = 0.2$)/Li$_{6.6}$Si$_{0.6}$Sb$_{0.4}$S$_5$I/Li-In cell exhibited a high initial discharge capacity of 190.6~mAh$\cdot$g$^{-1}$ and maintained a capacity retention of 68.3\% after 100 cycles.

\begin{figure*}
\centering
\includegraphics[width=0.92\textwidth]{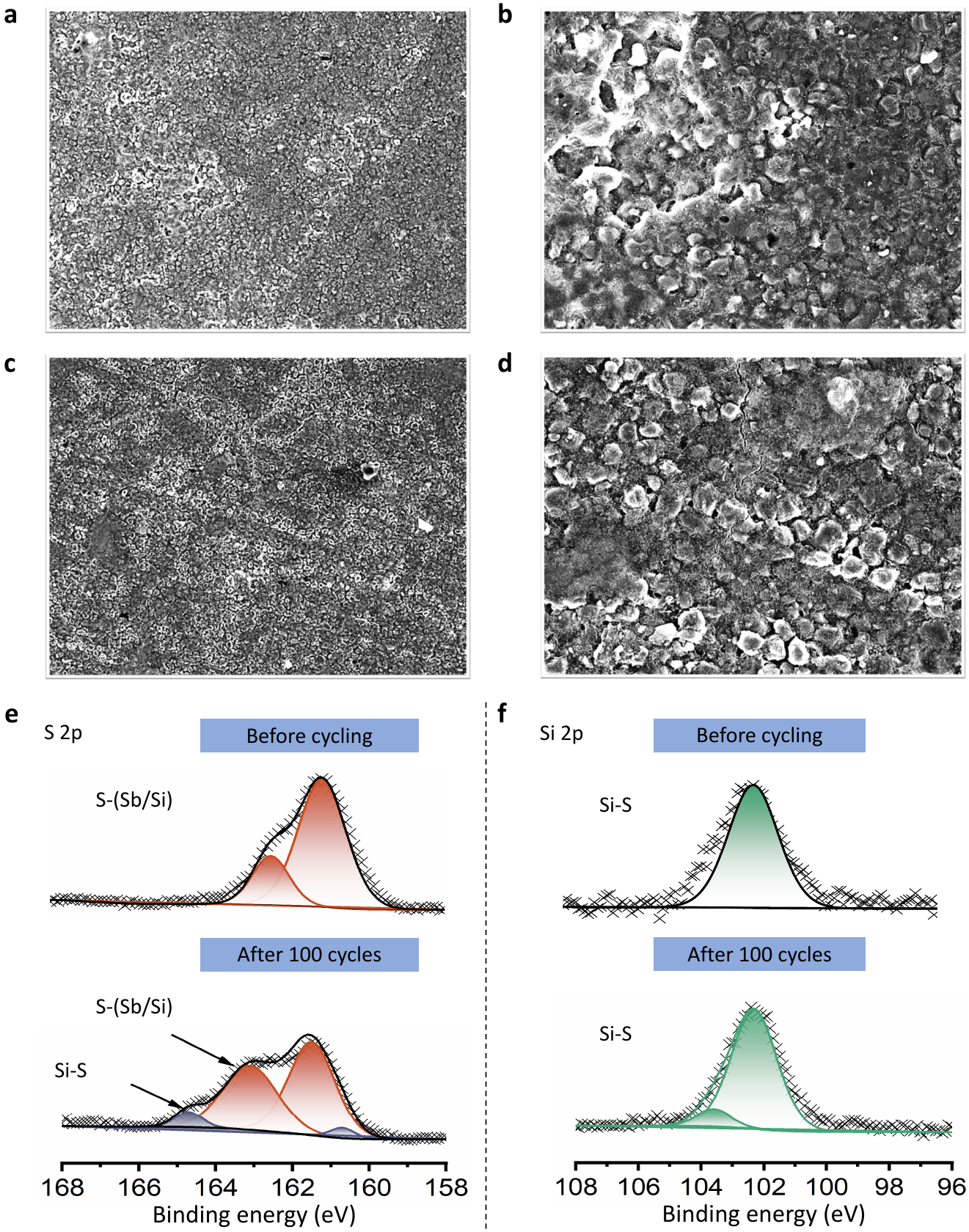}
\caption{Morphological and Chemical state evolutions of cathode interfaces. 
SEM images depicting the surface morphology of the LiNbO$_3$-coated cathode mixture with Li$_{6.6}$Si$_{0.6}$Sb$_{0.4}$S$_5$I (a,b) before and (c,d) after 100 charge/discharge cycles, demonstrating morphological evolution.
XPS spectra of (e) S 2p and (f) Si 2p from pristine and cycled LNO@NCM712/Li$_{6.6}$Si$_{0.6}$Sb$_{0.4}$S$_5$I/Li-In, demonstrating chemical state evolution upon cycling.
}
\label{Fig. 5}
\end{figure*}

Transitioning from surface morphology to structural stability, structural characterization of the electrode-electrolyte interface revealed well-integrated contact between the cathode particles and the Li$_{6.6}$Si$_{0.6}$Sb$_{0.4}$S$_5$I electrolyte in the pristine state (Fig.~\ref{Fig. 5}(a) and (b)). After 100 cycles at 0.2C (Fig.~\ref{Fig. 5}(c) and (d)), minor interfacial cracks were observed, attributed to repetitive volume changes during (de)lithiation. Despite these localized microcracks, the overall structural integrity was preserved, demonstrating robust interfacial compatibility throughout prolonged cycling. The composite cathode structure remained largely intact, with no evidence of severe degradation or particle disintegration. This structural preservation underscores the high interfacial compatibility between the optimized LNO@NCM cathode and the Li$_{6.6}$Si$_{0.6}$Sb$_{0.4}$S$_5$I electrolyte, highlighting the effectiveness of interface engineering in maintaining mechanical and electrochemical stability over extended cycling.

To complement the morphological and structural analyses, ex-situ XPS analysis was performed to further elucidate the dynamic interfacial evolution between the LNO@NCM cathode and the Li$_{6+x}$Si$_{x}$Sb$_{1-x}$S$_5$I electrolyte during cycling. The optimized cathode with 2 wt\% VGCF was analyzed before and after 100 cycles, as shown in Fig.~\ref{Fig. 5}(e) and (f), while the high-carbon cathode with 3 wt\% VGCF was further examined for comparison, as shown in Fig.~\ref{Fig. S9}. In the S 2\emph{p} spectra of the pristine interface, the main low-binding-energy components located at approximately 160.8--162.9 eV can be assigned to lattice S$^{2-}$ species in the original sulfide framework, including S-(Sb/Si) coordination and Si-S environments. For the optimized 2 wt\% VGCF cathode, the pristine interface displays characteristic S-(Sb/Si) coordination at 162.1/160.8 eV together with pre-existing Si-S bonds at 161.3/158.9 eV, indicating that only limited spontaneous interfacial reactions occur before electrochemical cycling.

After 100 cycles, the S 2\emph{p} spectra reveal clear chemical evolution at the cathode/electrolyte interface. For the optimized 2 wt\% VGCF cathode, the intensity of the S-(Sb/Si) component decreases by approximately 40\% and shifts positively by about 0.3 eV, suggesting partial sulfur oxidation and structural rearrangement during cycling. Meanwhile, the enhanced Si-S signal indicates the formation of electrolyte-derived interfacial species. In comparison, the 3 wt\% VGCF cathode exhibits more pronounced growth of high-binding-energy sulfur components at approximately 163.6-164.8 eV after cycling. These components are generally associated with S-(Sb/Si) environments, polysulfide-like species, elemental sulfur (S$^0$), and/or oxidized sulfur-containing species. Quantitative analysis shows that the high-binding-energy sulfur species account for approximately 58.8\% of the total S 2\emph{p} area in the 3 wt\% VGCF cathode, whereas the low-binding-energy sulfide components account for only approximately 41.2\%. This result indicates that excessive VGCF substantially accelerates sulfide-electrolyte decomposition and promotes the accumulation of sulfur-containing interfacial byproducts.

The Si 2\emph{p} spectra further confirm the participation of electrolyte-derived species in interphase formation. For the optimized 2 wt\% VGCF cathode, the pristine interface shows a sharp Si-S peak at 104.2 eV, which evolves into a dual-peak structure after cycling, with a new low-valence silicon species emerging at 102.9 eV and contributing approximately 25\% of the total Si 2\emph{p} intensity. This evolution suggests partial reconstruction of the local Si coordination environment and possible silicon-cluster formation. For the 3 wt\% VGCF cathode, the Si 2\emph{p} peaks after cycling located at approximately 102.6/103.2 eV can be assigned to Si-containing interphase species, such as Si-S and Si-O-S mixed environments. The peak shift, broadening, and intensity variation compared with the pristine state further indicate decomposition and reconstruction of the Li$_{6+x}$Si$_{x}$Sb$_{1-x}$S$_5$I framework at the cathode/electrolyte interface \cite{74}.

\begin{figure*}
\centering
\includegraphics[width=0.92\textwidth]{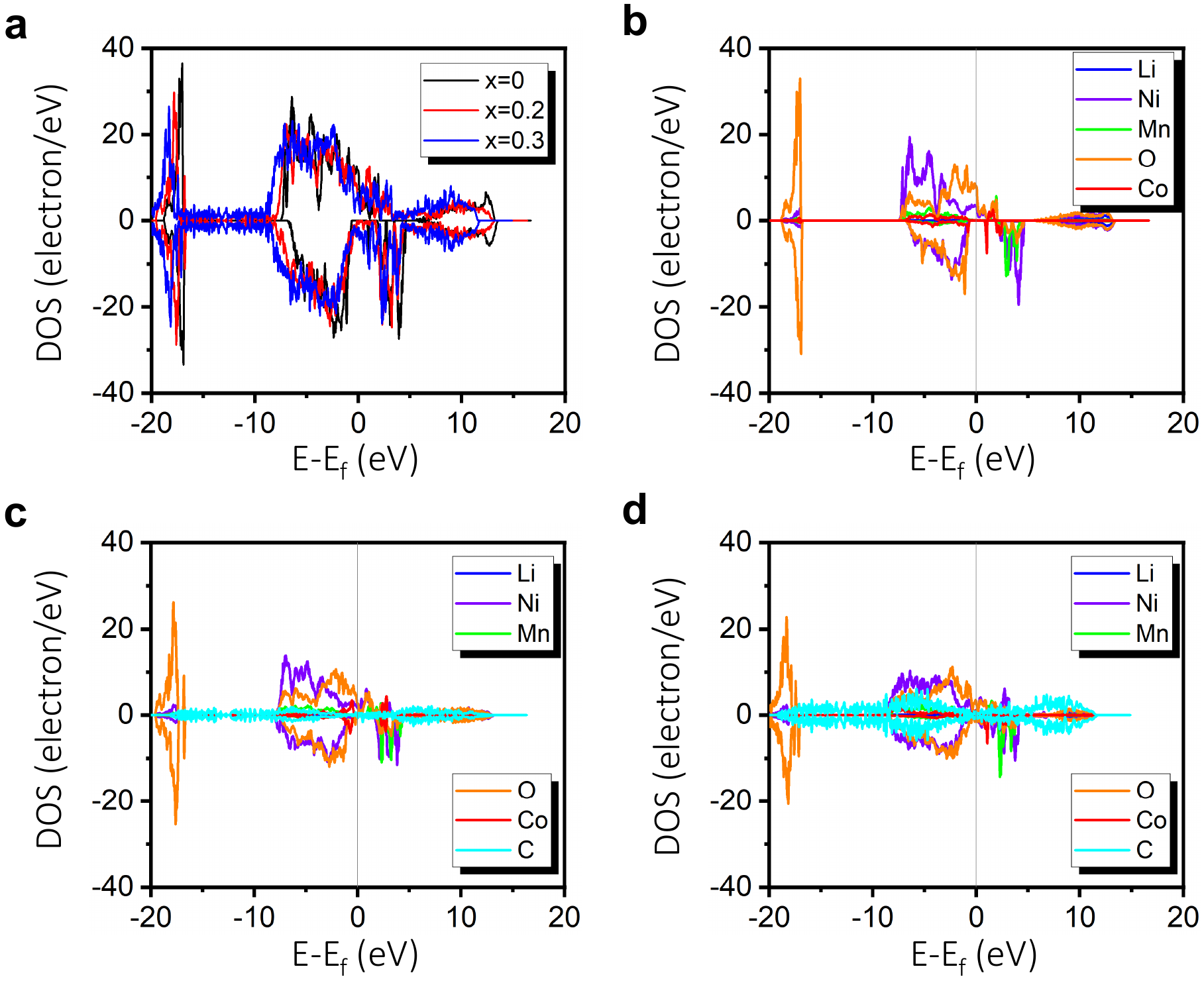}
\caption{Electronic properties of cathodes with varying carbon content. 
(a) Total density of states for cathodes with varied carbon amounts; partial density of states for batteries with carbon ratios of (b) $x = 0$, (c) $x = 0.2$, and (d) $x = 0.3$.
}
\label{Fig. 6}
\end{figure*}

To elucidate the microscopic mechanism by which conductive carbon (VGCF) content governs the performance of NCM-solid electrolyte composite cathodes, systematic DFT calculations were performed on three representative interface models: carbon-free, 2~wt\% VGCF, and 3~wt\% VGCF. The total density of states (DOS) in Fig.~\ref{Fig. 6}(a) shows that the DOS intensity at the Fermi level increases with carbon introduction, indicating enhanced electronic conductivity-a trend particularly pronounced in the spin-down channel. Partial DOS (PDOS) plots (Figs.~\ref{Fig. 6}(b)-(d)) further reveal that carbon incorporation markedly modifies the interfacial electronic states: in the carbon-free model, the PDOS peaks of Li, Ni, Mn, Co and O lie close to the Fermi level with high intensity, reflecting abundant yet highly localized states that lead to poor electronic conductivity and explain the observed low capacity but stable cycling; with 2~wt\% VGCF, the main peaks shift away from the Fermi level by about $0.3$~eV and their amplitudes decrease moderately, while a small carbon-derived peak emerges, indicating a balanced broadening of electronic bands that enhances electron delocalization without destabilizing the interface, corresponding to the optimal performance of high capacity and stable cycling; in the 3~wt\% VGCF model, the peaks shift only $\sim 0.15$~eV, their intensities drop more noticeably, and the carbon-related peak becomes much stronger, suggesting a highly delocalized, quasi-metallic interface that provides excellent initial conductivity (hence high initial capacity) but also acts as a continuous electron-leakage pathway that catalyses oxidative decomposition of the solid electrolyte, accounting for the severe capacity fading and voltage polarization observed at high carbon loading. Charge-density-difference maps (Fig.~\ref{Fig. S10}) illustrate the accompanying charge-transfer pathway, with the carbon layer acting as an electronic ``bridge'', and quantitative analysis shows the net electron gain of the carbon layer rises from 0.27 to 0.43 as carbon content increases from 0 to 0.3 (Tables~\ref{tab:S9}), confirming that higher carbon loading facilitates interfacial charge transfer. Thus, carbon content dictates the electrochemical performance by finely tuning the interfacial electronic structure, charge-transfer kinetics, and thermodynamic stability, establishing a clear ``structure-electronic-property-performance'' correlation that provides a theoretical foundation for optimizing conductive-additive design in solid-state batteries.

Overall, these complementary analyses reveal that the engineered electrode-electrolyte interface not only maintains structural integrity but also undergoes dynamic chemical evolution, collectively contributing to the enhanced cycling stability and electrochemical performance of the system.

\subsection{Wide-temperature adaptability and interfacial mechanisms of advanced solid-state batteries}

\begin{figure*}
\centering
\includegraphics[width=0.92\textwidth]{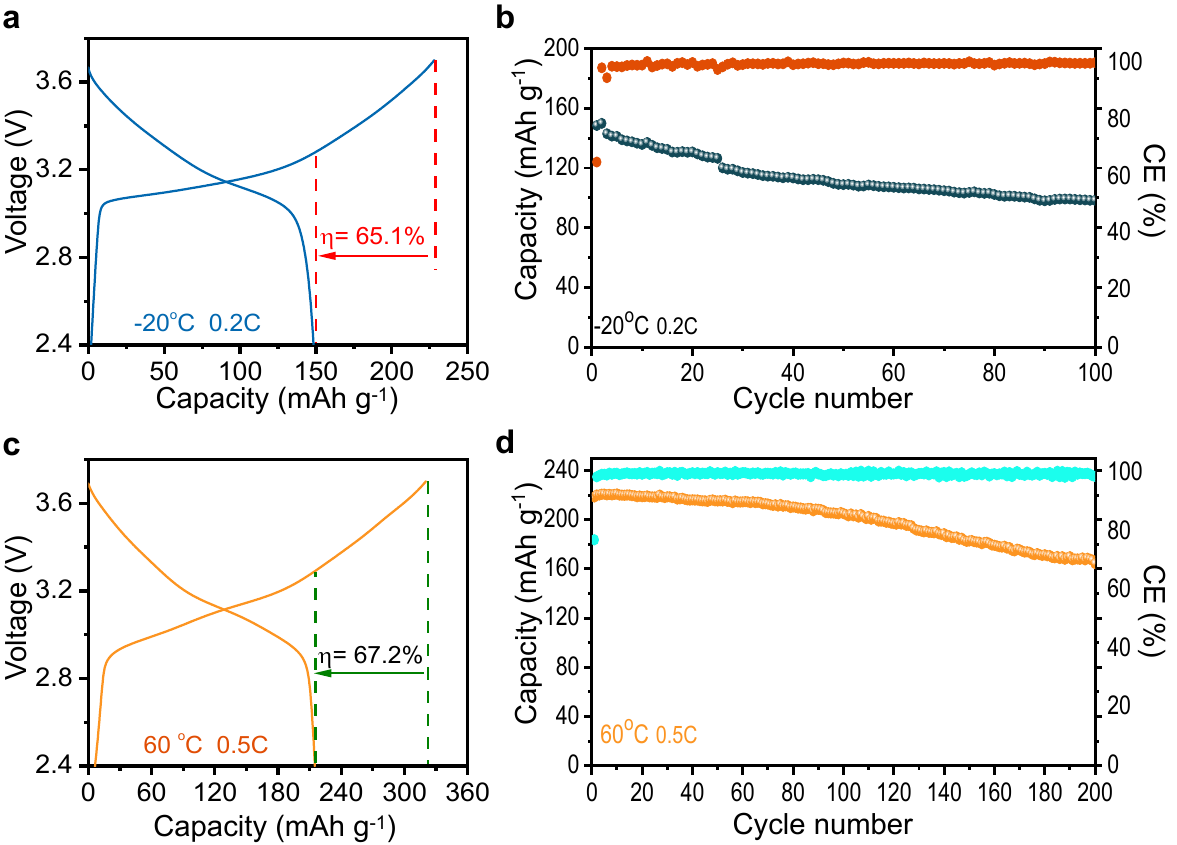}
\caption{Temperature-dependent cycling performance. 
Charge/discharge curves of LNO@NCM712/Li$_{6.6}$Si$_{0.6}$Sb$_{0.4}$S$_5$I/Li-In (a) at 0.2C and $-20~^\circ$C, with long-term cycling performance (b); (c) at 0.5C and $60~^\circ$C with long-term cycling performance (d), demonstrating stable electrochemical performance and exceptional durability under extreme temperature conditions.
}
\label{Fig. 7}
\end{figure*}

Subsequently, LNO@NCM712-VGCF ($x = 0.2$)/Li$_{6.6}$Si$_{0.6}$Sb$_{0.4}$S$_5$I/Li-In was evaluated under extreme temperature conditions. As illustrated in Fig.~\ref{Fig. 7}(a), the battery delivered an initial discharge capacity of 148.4~mAh~g$^{-1}$ and a Coulombic efficiency of 65.1\% at $-20~^\circ$C. After 100 cycles, it retained a capacity of 98.3~mAh~g$^{-1}$, corresponding to a capacity retention of 66.2\%. Although the initial Coulombic efficiency is relatively low, this behavior can be mainly attributed to the sluggish Li$^+$ transport kinetics in the solid electrolyte and composite cathode at low temperature, increased interfacial polarization, and incomplete utilization of the active material during the first cycle. Nevertheless, the cell still exhibits promising cyclability under low-temperature conditions, as shown in Fig.~\ref{Fig. 7}(b). The high-temperature performance was also investigated. As shown in Fig.~\ref{Fig. 7}(c), the LNO@NCM712-VGCF ($x = 0.2$)/Li$_{6.6}$Si$_{0.6}$Sb$_{0.4}$S$_5$I/Li-In cell exhibited an initial discharge capacity of 218.4~mAh~g$^{-1}$ and a Coulombic efficiency of 67.2\% at 0.5C under $60~^\circ$C. After cycling, the discharge capacity remained at 167.1~mAh~g$^{-1}$, corresponding to 76.5\% of its initial capacity. The relatively low initial Coulombic efficiency at elevated temperature may be associated with accelerated interfacial side reactions between the high-voltage NCM cathode and the sulfide electrolyte, as well as possible irreversible reactions at the Li-In/electrolyte interface. These irreversible processes can consume active lithium and contribute to the reduced initial Coulombic efficiency. Overall, these findings indicate that this battery architecture achieves robust electrochemical performance across both low- and high-temperature conditions, despite the interfacial and kinetic limitations during the initial cycle.

\begin{figure*}
\centering
\includegraphics[width=0.92\textwidth]{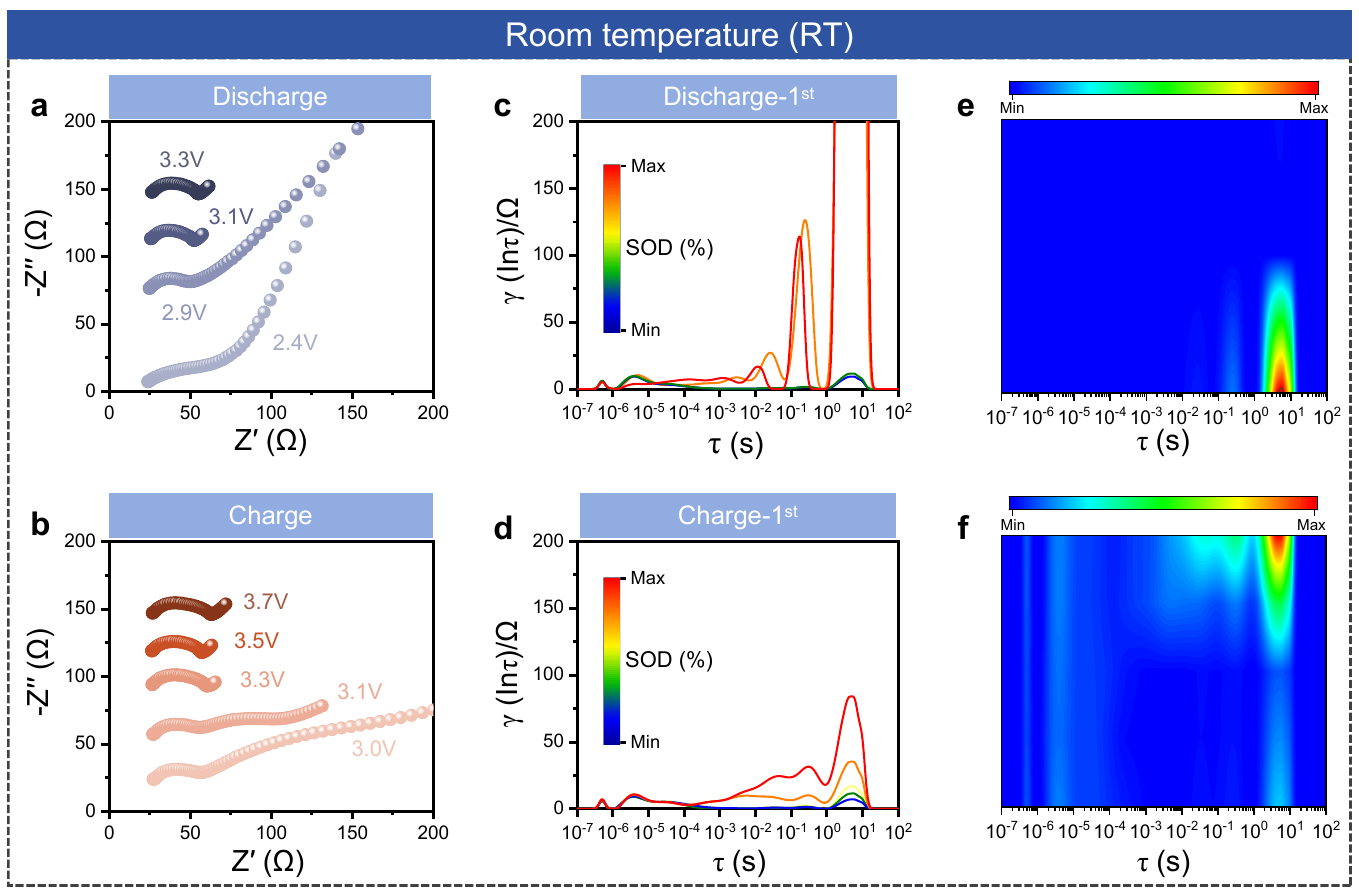}
\caption{Interfacial dynamics at room temperature. 
Electrochemical characterization of LNO@NCM712/Li$_{6.6}$Si$_{0.6}$Sb$_{0.4}$S$_5$I/Li-In batteries during the initial cycle. Impedance spectra under various (a,b) charge (SOC) and discharge (SOD) states at room temperature, with corresponding (c,d) DRT curves and (e,f) 2D intensity maps, elucidating dynamic interfacial processes.
}
\label{Fig. 8}
\end{figure*}

To investigate the dynamic interfacial processes in sulfide-based all-solid-state batteries at room temperature, in situ EIS was employed~\cite{75}. The typical Nyquist plots of the LNO@NCM712-VGCF (x = 0.2)/Li$_{6.6}$Si$_{0.6}$Sb$_{0.4}$S$_5$I/Li--In cell (Fig.~\ref{Fig. 8}(a) and (b)) display a depressed semicircle, attributed to charge-transfer mechanisms, and a linear region, corresponding to solid-state diffusion. During the initial discharge (Fig.~\ref{Fig. 8}(a)), the charge-transfer resistance increased from 61~$\Omega$ to 72~$\Omega$. In contrast, the charging process (Fig.~\ref{Fig. 8}(b)) exhibited a continuous decrease from 58~$\Omega$ to 49~$\Omega$, consistent with reversible electrode volume changes during (de)intercalation, as previously reported \cite{27}.

To further clarify the physical significance of the impedance response, distribution of relaxation times (DRT) analysis was conducted to deconvolute the EIS spectra into a series of relaxation processes with distinct characteristic time constants. Compared with conventional equivalent-circuit fitting, DRT analysis can more effectively distinguish overlapping electrochemical processes without assigning a specific circuit model in advance. In the DRT spectra, the peak position represents the characteristic relaxation time of a given process, while the peak intensity or integrated area reflects its relative resistance contribution. The DRT spectra (Fig.~\ref{Fig. 8}(c) and (d)) and corresponding 2D intensity maps (Fig.~\ref{Fig. 8}(e) and (f)) reveal multiple electrochemical phenomena during cycling. The high-frequency region of approximately $10^{-7}$--$10^{-6}$~s is mainly related to bulk and grain-boundary ion transport in the solid electrolyte. The intermediate relaxation region of approximately $10^{-5}$--$10^{-3}$~s is associated with the anode/electrolyte interfacial process, while the region of approximately $10^{-3}$--$10^{-2}$~s corresponds to ion transport and interfacial polarization at the cathode/electrolyte interface. The low-frequency region of approximately $10^{-2}$--$10^{1}$~s is mainly related to charge-transfer processes within the composite cathode and the Li--In alloy electrode \cite{76}.

Similar measurements performed at $-20~^{\circ}\mathrm{C}$ and $60~^{\circ}\mathrm{C}$ (Figs.~\ref{Fig. S11} and ~\ref{Fig. S12}) revealed consistent EIS and DRT trends, though with pronounced temperature dependence. At $-20~^{\circ}\mathrm{C}$, the total resistance increased significantly to approximately 295.6~$\Omega$ (Fig.~\ref{Fig. S11}(a) and (b)), accompanied by intensified DRT peaks (Fig.~\ref{Fig. S11}(c) and (d)). This indicates increased resistance contributions from sluggish Li$^{+}$ transport and enhanced interfacial polarization, which is consistent with the lower capacity and reduced initial Coulombic efficiency at low temperature. Conversely, the cell tested at $60~^{\circ}\mathrm{C}$ (Fig.~S12(a) and (b)) exhibited a low total resistance of approximately 15.2~$\Omega$ and subdued DRT features (Fig.~\ref{Fig. S12}(c) and (d)), reflecting reduced interfacial resistance, facilitated ion transport, and accelerated charge-transfer kinetics. These impedance characteristics are consistent with the observed electrochemical performance, demonstrating that DRT analysis provides a direct time-domain perspective on the evolution of ion transport and interfacial kinetics in the optimized all-solid-state cell.

\section{Conclusions}

In summary, we demonstrate an electrolyte--interface co-design strategy for wide-temperature sulfide-based all-solid-state lithium batteries. By introducing moderate Si$^{4+}$ substitution into the Sb-based iodide argyrodite framework, Li$_{6+x}$Si$_{x}$Sb$_{1-x}$S$_5$I was obtained with a high room-temperature ionic conductivity of $9.9 \times 10^{-3}$ S cm$^{-1}$ and a low activation energy of 0.18 eV. Structural and computational analyses indicate that the optimized Si content modifies the local lattice environment and lowers the Li-ion migration barrier, thereby promoting rapid Li-ion transport. When this electrolyte is integrated with a LiNbO$_{3}$@LiNi$_{0.7}$Co$_{0.1}$Mn$_{0.2}$O$_{2}$ composite cathode, controlled VGCF addition further balances electronic conduction and interfacial stability. In contrast to excessive carbon loading, which can facilitate electronic leakage and electrolyte decomposition, the optimized conductive-additive content enables efficient charge transport while suppressing interfacial degradation.

Benefiting from this coordinated electrolyte and cathode-interface design, the assembled LiNbO$_{3}$@LiNi$_{0.7}$Co$_{0.1}$Mn$_{0.2}$O$_{2}$ | Li$_{6+x}$Si$_{x}$Sb$_{1-x}$S$_5$I | Li--In all-solid-state battery delivers an initial discharge capacity of 174.8 mAh g$^{-1}$ and retains 68.2\% of its capacity after 300 cycles at 0.5C. The cell also maintains stable electrochemical performance over a wide temperature range from $-20~^\circ$C to $60~^\circ$C, demonstrating the applicability of this design under diverse thermal conditions. In situ EIS, DRT analysis, post-cycling microscopy, XPS characterization, and electronic-structure calculations collectively show that the enhanced cycling stability originates from the simultaneous improvement of bulk ion transport, interfacial contact, and electronically regulated cathode/electrolyte stability.

This study clarifies that electrolyte chemistry and composite-cathode architecture should be jointly optimized rather than independently adjusted in sulfide-based all-solid-state batteries. The results provide a practical design principle: high ionic conductivity must be paired with controlled interfacial electronic transport to avoid parasitic side reactions while maintaining sufficient reaction kinetics. Future work will focus on extending this strategy to more practical cell configurations, including thinner electrolyte layers, higher cathode areal loadings, lithium-metal anodes, and larger-format cells. Further operando characterization and long-term cycling under realistic pressure and temperature conditions will also be needed to clarify the dynamic evolution of sulfide-based interfaces and accelerate the practical deployment of wide-temperature all-solid-state batteries.

\clearpage

\section{Experimental Section}

\subsection{Controlled synthesis workflow}

Li$_{6+x}$Si$_{x}$Sb$_{1-x}$S$_5$I was synthesized from high-purity precursors: lithium sulfide (Li$_2$S, Sigma, 99.9\%), lithium iodide (LiI, Aladdin, 99.9\%), antimony pentasulfide (Sb$_2$S$_5$, Macklin, 99\%), silicon (Si, Aladdin, 99.9\%), and precipitated sulfur (S, Aladdin, 99.5\%). Reagents were weighed to stoichiometry and transferred into a tungsten carbide milling pot. Mechanical ball milling (Retsch PM 200) was conducted for 24~h at 500~rpm. The precursor powder was pelletized under 380~MPa, sealed in a quartz tube, and annealed at 450~$^\circ$C for 6~h. All operations were carried out in an argon-filled glovebox (H$_2$O $<$ 0.1~ppm, O$_2$ $<$ 0.1~ppm), with air-free transfers to the furnace. The synthesized Li$_{6+x}$Si$_{x}$Sb$_{1-x}$S$_5$I was subsequently used for composite cathode fabrication and further characterization.

\subsection{Uniform cathode mixing}

Composite cathodes comprising LiNbO$_3$-coated LiNi$_{0.7}$Co$_{0.1}$Mn$_{0.2}$O$_2$, Li$_{6.6}$Si$_{0.6}$Sb$_{0.4}$S$_5$I, and vapor-grown carbon fibers (VGCF) were prepared by ball milling. Mixtures with weight ratios of NCM712:SEs:VGCF = 7:3:0, 7:2.8:0.2, and 7:2.7:0.3 were milled in a ZrO$_2$ pot at 180~rpm for 1~h to achieve uniform dispersion. These cathodes were then used to evaluate interfacial compatibility and electronic percolation.

\subsection{Correlative structure–chemistry analysis}

Phase identification was performed by X-ray diffraction (XRD, Cu K$\alpha$, Smart Lab-SE Powder) over 10--80$^\circ$ in 2$\theta$, and Rietveld refinement was conducted using GSAS to assess phase purity and lattice parameters. Raman spectroscopy (Lab EAM HR800, 532~nm) was collected over 100–800~cm$^{-1}$ to probe polyanion vibrations. Morphology and elemental distribution were examined by field-emission scanning electron microscopy (SEM, Nova NanoSEM 450) with energy dispersive spectroscopy (EDS) mapping, and surface chemistry was assessed by X-ray photoelectron spectroscopy (XPS, AXIS-ULTRA DLD-600W, Shimadzu-Kratos). The binding energies were calibrated with reference to the C 1\emph{s} peak at 284.8 eV. The Si 2\emph{p} and S 2\emph{p} spectra were fitted after background subtraction, and the peak-shape parameters were refined according to the actual spectral features and assigned chemical states. During the fitting of the Si 2\emph{p} spectra, the FWHM values of the Si-related components were constrained within a physically reasonable range, with a representative Si 2\emph{p} component centered at approximately 101.70 eV and an FWHM of approximately 2.38 eV. In the S 2\emph{p} spectra, the low-binding-energy components located at approximately 160.8--162.9 eV were assigned to sulfide-related species. These results informed selection of voltage windows, current densities, and coating strategies for electrochemical testing.

\subsection{Transport properties and stability window}

Electrochemical impedance spectroscopy (EIS) was performed on a Bio-Logic analyzer from 0.1 Hz to 2 MHz with an AC amplitude of 50 mV over a temperature range of 30--70~$^\circ$C. For ionic conductivity measurements, 100 mg of Li$_{6+x}$Si$_{x}$Sb$_{1-x}$S$_5$I powder was cold-pressed into a pellet with a diameter of $\sim$10 mm and a thickness of 0.069 cm under 380 MPa. The pellet area was 0.785 cm$^2$, and the apparent density was calculated from the pellet mass and volume ($\sim$1.85 g cm$^{-3}$). The relative density was further obtained by comparing the apparent density with the theoretical density derived from the refined crystal structure.

For EIS measurements, the electrolyte pellet was sandwiched between two stainless-steel current collectors as ion-blocking electrodes to form a symmetric SS $\vert$ electrolyte $\vert$ SS cell. The impedance spectra were fitted using the equivalent-circuit model $R_s$-($R_{\mathrm{SE}} \parallel \mathrm{CPE}_{\mathrm{SE}}$)-$\mathrm{CPE}_{\mathrm{SS}}$, where $R_s$ represents the series resistance, $R_{\mathrm{SE}}$ is the resistance associated with ion transport through the solid electrolyte pellet, $\mathrm{CPE}_{\mathrm{SE}}$ accounts for the non-ideal capacitive response of the electrolyte, and $\mathrm{CPE}_{\mathrm{SS}}$ represents the low-frequency blocking-electrode polarization from the stainless-steel current collectors.

The resistance used for the ionic conductivity calculation was obtained from the fitted $R_{\mathrm{SE}}$ value using the CHI760E software. Since the bulk and grain-boundary contributions could not be unambiguously separated for all samples, especially for spectra without a clearly resolved semicircle, $R_{\mathrm{SE}}$ was regarded as the effective total ion-transport resistance of the solid electrolyte pellet. Therefore, $R_{\mathrm{SE}}$ was used for the ionic conductivity and activation energy calculations to ensure a consistent comparison among different samples.

The ionic conductivity was calculated according to:

\begin{equation}
\sigma = \frac{L}{R \times A}
\end{equation}

where $\sigma$ is the ionic conductivity, $L$ is the pellet thickness, $R$ is the fitted resistance, and $A$ is the effective electrode area. In this work, $R$ corresponds to $R_{\mathrm{SE}}$ obtained from the equivalent-circuit fitting.

The activation energy was calculated using the Arrhenius relationship:

\begin{equation}
\sigma T = A_0 \exp\left(-\frac{E_a}{k_B T}\right)
\end{equation}

or equivalently:

\begin{equation}
\ln(\sigma T) = \ln A_0 - \frac{E_a}{k_B T}
\end{equation}

where $E_a$ is the activation energy, $k_B$ is the Boltzmann constant, $T$ is the absolute temperature, and $A_0$ is the pre-exponential factor. The activation energy was obtained from the slope of the linear fitting of $\ln(\sigma T)$ versus $1000/T$.

Chronoamperometry (CA) measurements were conducted to determine the lithium-ion transference number of the Li$_{6+x}$Si$_{x}$Sb$_{1-x}$S$_5$I electrolytes. Symmetric cells were polarized under a constant voltage of 10 mV. The initial current ($I_0$) and steady-state current ($I*{\mathrm{ss}}$) were obtained from the current--time profiles. EIS measurements were performed before and after polarization to determine the initial resistance ($R_0$) and steady-state resistance ($R*{\mathrm{ss}}$), respectively. The lithium-ion transference number ($t_+$) was calculated according to:
\[
t_+ =
\frac{
I_0\left(\Delta V - I_{\mathrm{ss}}R_{\mathrm{ss}}\right)
}{
I_{\mathrm{ss}}\left(\Delta V - I_0R_0\right)
}
\]
where $\Delta V$ is the applied polarization voltage of 10 mV, $I_0$ is the initial current, $I_{\mathrm{ss}}$ is the steady-state current, $R_0$ is the initial resistance, and $R_{\mathrm{ss}}$ is the steady-state resistance.

The electrochemical stability window was determined using asymmetric \(\mathrm{Li_{6+x}Si_xSb_{1-x}S_5I@C}\allowbreak/\allowbreak\mathrm{Li_{6+x}Si_xSb_{1-x}S_5I}\)/Li--In cells on a Solartron 1260, scanned from 0 to 4 V vs. In/Li--In at 0.1 mV s$^{-1}$. For cathode preparation, Li$_{6+x}$Si$_{x}$Sb$_{1-x}$S$_{5}$I and VGCF were mixed at a mass ratio of 7:3 and ball-milled for 1 h at 180 rpm, and a Li--In alloy was placed on the opposite side of the SE pellet. These electrochemical data guided the assembly and evaluation of all-solid-state lithium batteries.

\subsection{Full-cell construction and diagnostics}
First, 100~mg of Li$_{6.6}$Si$_{0.6}$Sb$_{0.4}$S$_5$I was pressed into $\sim$10~mm-diameter pellets in a mold cell under 10.0~MPa. Then, 5.0~mg of composite cathode was uniformly pressed onto one side under 380.0~MPa, and a Li-In alloy anode was placed on the other side under 150.0~MPa. Cycling performance of ASSLBs was evaluated by galvanostatic charge/discharge at ambient temperature from 2.4 to 3.7~V (vs. Li-In) on a Neware CT4008 at various current densities. In-situ and ex-situ EIS were performed on a Zahner instrument from $10^{-1}$ to $10^{7}$~Hz with a perturbation amplitude of 10~mV; in-situ spectra were collected at selected states of charge, and ex-situ measurements were taken after defined cycle counts. This protocol enabled a systematic assessment of ASSLB performance while maintaining air-free handling throughout fabrication and testing.

\subsection{Computational methods}

First-principles calculations were performed using the Vienna Ab initio Simulation Package (VASP, version 5.4.4) within the projector augmented wave (PAW) framework. The exchange--correlation interactions were described using the generalized gradient approximation (GGA) with the Perdew-Burke-Ernzerhof (PBE) functional.

For the Li-ion migration calculations in Li$_{6+x}$Si$_{x}$Sb$_{1-x}$S$_5$I electrolytes, spin polarization was not considered because no transition-metal magnetic elements are present in these argyrodite systems. A plane-wave kinetic energy cutoff of 450 eV was used, and the Brillouin zone was sampled using a $5 \times 5 \times 5$ Monkhorst--Pack $k$-point grid. Structural relaxations were performed using the conjugate-gradient algorithm until the residual forces on each atom were less than 0.01 eV \AA$^{-1}$, and the self-consistent-field convergence criterion was set to $5 \times 10^{-5}$ eV. The climbing-image nudged elastic band (CI-NEB) method was employed to determine the Li$^+$ migration energy barriers. Eight intermediate images were inserted between the initial and final states, and a spring constant of 5.0 eV \AA$^{-2}$ was used. The transition states were verified by the presence of a single imaginary frequency along the reaction coordinate. The pristine argyrodite-type Li$_{6}$SbS$_5$I structure was adopted from the work of Yi et al.~\cite{77} in which the lattice parameters and atomic positions were validated against experimental X-ray diffraction data. This structure was used as the parent framework for constructing the Si-substituted models. The Li$_{6.6}$Si$_{0.6}$Sb$_{0.4}$S$_5$I and Li$_{6.75}$Si$_{0.75}$Sb$_{0.25}$S$_5$I models were generated by partially substituting Sb atoms with Si atoms at the cation sites while maintaining charge neutrality. The selected Si contents were consistent with the experimentally reported compositions in Yi et al.'s work, enabling direct comparison with the observed ionic conductivity trends. For each composition, several possible Si distributions on the Sb sublattice were considered, and the most energetically favorable configuration was selected after full structural relaxation using DFT. These models were used to systematically investigate how increasing Si substitution modulates the charge distribution and Li-ion migration behavior in Li$_{6+x}$Si$_{x}$Sb$_{1-x}$S$_5$I electrolytes.

For the NCM/graphene interface models, spin-polarized calculations were performed because transition-metal elements are present in LiNi$_{0.7}$Co$_{0.1}$Mn$_{0.2}$O$_{2}$. The DFT-D3 dispersion correction was applied to describe van der Waals interactions between the NCM surface and carbon layers. A plane-wave cutoff energy of 480 eV was used. The electronic convergence criterion was set to $1 \times 10^{-5}$ eV, and all structures were fully relaxed until the residual forces on each atom were less than 0.02 eV \AA$^{-1}$. Gaussian smearing with a width of 0.03 eV was employed. For the transition-metal-containing NCM models, the DFT+$U$ method was applied to better describe the localized 3$d$ electrons, with $U-J$ values of 6.2, 3.9, and 3.32 eV for Ni, Mn, and Co, respectively.

For the NCM/graphene interface models, the NCM structure was first constructed from a LiNi$_{0.7}$Co$_{0.1}$Mn$_{0.2}$O$_{2}$ primitive cell containing 12 atoms, with initial lattice parameters of $a=b=2.87$ \AA{}, $c=14.23$ \AA{}, $\alpha=\beta=90^\circ$, and $\gamma=120^\circ$. The primitive cell was expanded into a $2 \times 2 \times 1$ supercell, and cation substitution was performed with an atomic ratio of Li:Ni:Mn:Co = 11:10:2:1 to obtain an NCM supercell containing 48 atoms. After full lattice optimization, the optimized parameters were $a=b=5.74$ \AA{}, $c=14.16$ \AA{}, $\alpha=90^\circ$, $\beta=90.18^\circ$, and $\gamma=120^\circ$.

The carbon component was represented by graphene layers because VGCF possesses a graphitic carbon framework. A single-layer graphene model containing 8 carbon atoms was constructed with lattice parameters of $a=b=4.92$ \AA{}, $\alpha=\beta=90^\circ$, and $\gamma=120^\circ$. This graphene layer was placed on the NCM (001) surface to construct the NCM-1 heterostructure model. The initial in-plane lattice parameter of the interface was set as the average value of the optimized NCM surface cell and graphene sheet, namely $a=b=5.33$ \AA{}. This introduced a tensile strain of 8.33\% in graphene and a compressive strain of 7.14\% in the NCM surface, both below 10\%. The interface model was then fully optimized, including both lattice parameters and atomic positions. The optimized NCM-1 model contained 56 atoms, with final lattice parameters of $a=5.48$ \AA{}, $b=5.43$ \AA{}, $c=17.99$ \AA{}, $\alpha=89.81^\circ$, $\beta=89.38^\circ$, and $\gamma=119.73^\circ$. Following the same procedure, the NCM-3 model was constructed by increasing the graphene layer number to three, resulting in a 72-atom model with optimized lattice parameters of $a=b=5.23$ \AA{}, $c=25.34$ \AA{}, $\alpha=90.24^\circ$, $\beta=90.40^\circ$, and $\gamma=119.81^\circ$.

For structural optimization of the NCM/graphene models, the Brillouin zone was sampled using a $5 \times 5 \times 2$ Monkhorst--Pack $k$-point mesh centered at the Gamma point. For each optimized model, a single-point energy calculation was first performed with the same $5 \times 5 \times 2$ $k$-point mesh to obtain the converged charge density file. Density of states (DOS) calculations were then conducted using a denser $10 \times 10 \times 4$ $k$-point mesh with ICHARG = 11.

Charge-density-difference analysis was conducted to visualize interfacial charge redistribution between NCM and graphene. The charge density difference was calculated according to:
\[
\Delta \rho = \rho_{\mathrm{AB}} - \rho_{\mathrm{A}} - \rho_{\mathrm{B}},
\]
where $\rho_{\mathrm{AB}}$ is the charge density of the optimized NCM/graphene interface, and $\rho_{\mathrm{A}}$ and $\rho_{\mathrm{B}}$ are the charge densities of the isolated NCM slab and graphene layer in the same atomic positions, respectively. The charge density difference was visualized using VESTA. In the interface model, the NCM phase is located at the bottom and the graphene carbon layer is located at the top, corresponding to the NCM (001) surface.

\clearpage

\section*{CRediT authorship contribution statement}

\textbf{Liang Ming:} Conceptualization, data curation, formal analysis, investigation, visualization, writing-original draft. 
\textbf{Qizhiran Sun:} Data curation, formal analysis, investigation, visualization. 
\textbf{Guanping Xu:} Data curation, formal analysis, investigation, visualization. 
\textbf{Muqing Su:} Data curation, formal analysis, investigation, visualization. 
\textbf{Enyan Zhao:} Data curation, formal analysis, investigation, visualization. 
\textbf{Wenzhe Gu:} Data curation, formal analysis, investigation, visualization. 
\textbf{Weng-Fu Io:} Data curation, formal analysis, investigation, visualization. 
\textbf{Kwun Nam Hui:} Conceptualization, methodology, supervision, visualization, writing-review \& editing.
\textbf{Chuang Yu:} Conceptualization, funding acquisition, methodology, project administration, supervision, visualization, writing-review \& editing.
\textbf{Hai-Feng Li:} Conceptualization, funding acquisition, methodology, project administration, supervision, visualization, writing-review \& editing.

\section*{Declaration of Competing Interest}

The authors declare that they have no known competing financial interests or personal relationships that could have appeared to influence the work reported in this paper.

\section*{Acknowledgments}

This work was partially supported by the National Key Research and Development Program (2021YFB2400300) and National Natural Science Foundation of China (Nos. 52177214). 
The work at the University of Macau was supported by the Science and Technology Development Fund, Macao SAR (File Nos. 0104/2024/AFJ, 0002/2024/TFP, and 0115/2024/RIB2), University of Macau (MYRG-GRG2024-00158-IAPME and MYRG-GRG2025-00251-IAPME), and the Guangdong-Hong Kong-Macao Joint Laboratory for Neutron Scattering Science and Technology (Grant No. 2019B121205003). 

\section*{Data Availability}

Data will be made available on request.

\clearpage

\clearpage

\clearpage

\clearpage

\clearpage

\clearpage

\clearpage

\clearpage

\clearpage

\bibliographystyle{elsarticle-num-names}
\bibliography{MingLiang-V3}

@article{1,
  title={Interfaces and interphases in all-solid-state batteries with inorganic solid electrolytes},
  author={Banerjee, Abhik and Wang, Xuefeng and Fang, Chengcheng and Wu, Erik A and Meng, Ying Shirley},
  journal={Chemical Reviews},
  volume={120},
  number={14},
  pages={6878--6933},
  year={2020},
  publisher={ACS Publications}
}

@article{2,
  title={A review of challenges and issues concerning interfaces for all-solid-state batteries},
  author={Lim, Hee-Dae and Park, Jae-Ho and Shin, Hyeon-Ji and Jeong, Jiwon and Kim, Jun Tae and Nam, Kyung-Wan and Jung, Hun-Gi and Chung, Kyung Yoon},
  journal={Energy Storage Materials},
  volume={25},
  pages={224--250},
  year={2020},
  publisher={Elsevier}
}

@article{3,
  title={Sulfide-based solid-state electrolytes: synthesis, stability, and potential for all-solid-state batteries},
  author={Zhang, Qing and Cao, Daxian and Ma, Yi and Natan, Avi and Aurora, Peter and Zhu, Hongli},
  journal={Advanced Materials},
  volume={31},
  number={44},
  pages={1901131},
  year={2019},
  publisher={Wiley Online Library}
}

@article{4,
  title={From nanoscale interface characterization to sustainable energy storage using all-solid-state batteries},
  author={Tan, Darren HS and Banerjee, Abhik and Chen, Zheng and Meng, Ying Shirley},
  journal={Nature Nanotechnology},
  volume={15},
  number={3},
  pages={170--180},
  year={2020},
  publisher={Nature Publishing Group UK London}
}

@article{5,
  title={High-power all-solid-state batteries using sulfide superionic conductors},
  author={Kato, Yuki and Hori, Satoshi and Saito, Toshiya and Suzuki, Kota and Hirayama, Masaaki and Mitsui, Akio and Yonemura, Masao and Iba, Hideki and Kanno, Ryoji},
  journal={Nature Energy},
  volume={1},
  number={4},
  pages={1--7},
  year={2016},
  publisher={Nature Publishing Group}
}

@article{6,
title = {Predicting doping strategies for ternary nickel–cobalt–manganese cathode materials to enhance battery performance using graph neural networks},
author = {Zirui Zhao and Dong Luo and Shuxing Wu and Kaitong Sun and Zhan Lin and Hai-Feng Li},
journal = {Journal of Energy Storage},
volume = {98},
pages = {112982},
year = {2024}
}

@article{7,
author = {Wang, Xiaoke and Hu, Sanlue and Xiao, Fangyuan and Xu, Guanping and Zhao, Zirui and Chang, Caiyun and Zhang, Xiangyong and Han, Cuiping and Li, Hai-Feng and Cheng, Hui-Ming},
title = "{A Highly Water-Retentive Electrolyte Enabling Stable Aqueous Zinc-Ion Batteries Across a Wide Temperature Range}",
journal = {Advanced Functional Materials},
volume = {},
number = {},
pages = {e19785},
year = {2025}
}

@article{8,
  title={Issues and advances in scaling up sulfide-based all-solid-state batteries},
  author={Lee, Jieun and Lee, Taegeun and Char, Kookheon and Kim, Ki Jae and Choi, Jang Wook},
  journal={Accounts of Chemical Research},
  volume={54},
  number={17},
  pages={3390--3402},
  year={2021},
  publisher={ACS Publications}
}

@article{9,
  title={Interface issues and challenges in all-solid-state batteries: lithium, sodium, and beyond},
  author={Lou, Shuaifeng and Zhang, Fang and Fu, Chuankai and Chen, Ming and Ma, Yulin and Yin, Geping and Wang, Jiajun},
  journal={Advanced Materials},
  volume={33},
  number={6},
  pages={2000721},
  year={2021},
  publisher={Wiley Online Library}
}

@article{10,
  title={Interfacial challenges for all-solid-state batteries based on sulfide solid electrolytes},
  author={Wang, Shuo and Fang, Ruyi and Li, Yutao and Liu, Yuan and Xin, Chengzhou and Richter, Felix H and Nan, Ce-Wen},
  journal={Journal of Materiomics},
  volume={7},
  number={2},
  pages={209--218},
  year={2021},
  publisher={Elsevier}
}

@article{11,
  title={Understanding chemical stability issues between different solid electrolytes in all-solid-state batteries},
  author={Riphaus, Nathalie and Stiaszny, Barbara and Beyer, Hans and Indris, Sylvio and Gasteiger, Hubert A and Sedlmaier, Stefan J},
  journal={Journal of The Electrochemical Society},
  volume={166},
  number={6},
  pages={A975--A983},
  year={2019},
  publisher={The Electrochemical Society}
}

@article{12,
author = {Zhao, Zirui and Xia, Junchao and Wu, Si and Wang, Xiaoke and Xu, Guanping and Zhu, Yinghao and Sun, Jing and Li, Hai-Feng},
title = "{Insights Into Dendritic Growth Mechanisms in Batteries: A Combined Machine Learning and Computational Study}",
journal = {Battery Energy},
volume = {4},
number = {5},
pages = {e70015},
year = {2025}
}

@Article{13,
author={Zhao, Zirui and Wang, Xiaoke and Wu, Si and Zhou, Pengfei and Zhao, Qian and Xu, Guanping and Sun, Kaitong and Li, Hai-Feng},
title="{Deep learning-driven evaluation and prediction of ion-doped NASICON materials for enhanced solid-state battery performance}",
journal={AAPPS Bulletin},
year={2024},
volume={34},
number={1},
pages={26}
}

@article{14,
author = {Zhou, Pengfei and Zhao, Zirui and Sun, Kaitong and Zhao, Qian and Xiao, Fangyuan and Fu, Ying and Li, Hai-Feng},
title = "{Machine Learning Guided Cobalt-doping Strategy for Solid-state NASICON Electrolytes}",
journal = {European Journal of Inorganic Chemistry},
volume = {26},
number = {26},
pages = {e202300382},
year = {2023}
}

@article{15,
  title={Lithium/sulfide all-solid-state batteries using sulfide electrolytes},
  author={Wu, Jinghua and Liu, Sufu and Han, Fudong and Yao, Xiayin and Wang, Chunsheng},
  journal={Advanced Materials},
  volume={33},
  number={6},
  pages={2000751},
  year={2021},
  publisher={Wiley Online Library}
}

@article{16,
  title={Sulfide-based solid-state electrolytes: synthesis, stability, and potential for all-solid-state batteries},
  author={Zhang, Qing and Cao, Daxian and Ma, Yi and Natan, Avi and Aurora, Peter and Zhu, Hongli},
  journal={Advanced Materials},
  volume={31},
  number={44},
  pages={1901131},
  year={2019},
  publisher={Wiley Online Library}
}

@article{17,
  title={Design strategies, practical considerations, and new solution processes of sulfide solid electrolytes for all-solid-state batteries},
  author={Park, Kern Ho and Bai, Qiang and Kim, Dong Hyeon and Oh, Dae Yang and Zhu, Yizhou and Mo, Yifei and Jung, Yoon Seok},
  journal={Advanced Energy Materials},
  volume={8},
  number={18},
  pages={1800035},
  year={2018},
  publisher={Wiley Online Library}
}

@article{18,
  title={Issues and advances in scaling up sulfide-based all-solid-state batteries},
  author={Lee, Jieun and Lee, Taegeun and Char, Kookheon and Kim, Ki Jae and Choi, Jang Wook},
  journal={Accounts of Chemical Research},
  volume={54},
  number={17},
  pages={3390--3402},
  year={2021},
  publisher={ACS Publications}
}

@article{19,
  title={Challenges, interface engineering, and processing strategies toward practical sulfide-based all-solid-state lithium batteries},
  author={Liang, Yuhao and Liu, Hong and Wang, Guoxu and Wang, Chao and Ni, Yu and Nan, Ce-Wen and Fan, Li-Zhen},
  journal={InfoMat},
  volume={4},
  number={5},
  pages={e12292},
  year={2022},
  publisher={Wiley Online Library}
}

@article{20,
  title={Issues and challenges for bulk-type all-solid-state rechargeable lithium batteries using sulfide solid electrolytes},
  author={Jung, Yoon Seok and Oh, Dae Yang and Nam, Young Jin and Park, Kern Ho},
  journal={Israel Journal of Chemistry},
  volume={55},
  number={5},
  pages={472--485},
  year={2015},
  publisher={Wiley Online Library}
}

@article{21,
  title={Innovative approaches to {Li}-argyrodite solid electrolytes for all-solid-state lithium batteries},
  author={Zhou, Laidong and Minafra, Nicolo and Zeier, Wolfgang G and Nazar, Linda F},
  journal={Accounts of Chemical Research},
  volume={54},
  number={12},
  pages={2717--2728},
  year={2021},
  publisher={ACS Publications}
}

@article{22,
  title={Research progress in {Li}-argyrodite-based solid-state electrolytes},
  author={Bai, Xiangtao and Duan, Yi and Zhuang, Weidong and Yang, Rong and Wang, Jiantao},
  journal={Journal of Materials Chemistry A},
  volume={8},
  number={48},
  pages={25663--25686},
  year={2020},
  publisher={Royal Society of Chemistry}
}

@article{23,
  title={Studies of lithium argyrodite solid electrolytes for all-solid-state batteries},
  author={Rao, R Prasada and Adams, S},
  journal={Physica Status Solidi (a)},
  volume={208},
  number={8},
  pages={1804--1807},
  year={2011},
  publisher={Wiley Online Library}
}

@article{24,
  title={Lithium argyrodite sulfide electrolytes with high ionic conductivity and air stability for all-solid-state {Li}-ion batteries},
  author={Lee, Yongheum and Jeong, Jiwon and Lee, Ho Jun and Kim, Mingony and Han, Daseul and Kim, Hyoungchul and Yuk, Jong Min and Nam, Kyung-Wan and Chung, Kyung Yoon and Jung, Hun-Gi and others},
  journal={ACS Energy Letters},
  volume={7},
  number={1},
  pages={171--179},
  year={2021},
  publisher={ACS Publications}
}

@article{25,
  title={Recent progress on lithium argyrodite solid-state electrolytes},
  author={Peng, Linfeng and Yu, Chuang and Wei, Chaochao and Liao, Cong and Chen, Shuai and Zhang, Long and Cheng, Shijie and Xie, Jia},
  journal={Acta Phys. Chim. Sin},
  volume={39},
  number={7},
  pages={2211034},
  year={2023}
}

@article{26,
  title={Understanding the origin of enhanced {Li}-ion transport in nanocrystalline argyrodite-type {Li}$_{6}${PS}$_{5}${I}},
  author={Brinek, Marina and Hiebl, Caroline and Wilkening, H Martin R},
  journal={Chemistry of Materials},
  volume={32},
  number={11},
  pages={4754--4766},
  year={2020},
  publisher={ACS Publications}
}

@article{27,
  title={With a little help from {\(^{31}\){P}} {NMR}{:} the complete picture on localized and long-range {Li}\(^{+}\) diffusion in {Li}$_{6}${PS}$_{5}${I}},
  author={Hogrefe, Katharina and Hanghofer, Isabel and Wilkening, H Martin R},
  journal={The Journal of Physical Chemistry C},
  volume={125},
  number={41},
  pages={22457--22463},
  year={2021},
  publisher={ACS Publications}
}

@article{28,
  title={Synthetic optimization and application of {Li}-argyrodite {Li}$_{6}${PS}$_{5}${I} in solid-state battery at different temperatures},
  author={He, Zhen-Yuan and Zhang, Zi-Qi and Yu, Ming and Yu, Chuang and Ren, Hao-Tian and Zhang, Jun-Zhao and Peng, Lin-Feng and Zhang, Long and Cheng, Shi-Jie and Xie, Jia},
  journal={Rare Metals},
  volume={41},
  number={3},
  pages={798--805},
  year={2022},
  publisher={Springer}
}

@article{29,
  title={Silicon-doped argyrodite solid electrolyte {Li}$_{6}${PS}$_{5}${I} with improved ionic conductivity and interfacial compatibility for high-performance all-solid-state lithium batteries},
  author={Zhang, Jun and Li, Lujie and Zheng, Chao and Xia, Yang and Gan, Yongping and Huang, Hui and Liang, Chu and He, Xinping and Tao, Xinyong and Zhang, Wenkui},
  journal={ACS Applied Materials \& Interfaces},
  volume={12},
  number={37},
  pages={41538--41545},
  year={2020},
  publisher={ACS Publications}
}

@article{30,
  title={Further evidence for energy landscape flattening in the superionic argyrodites {Li}$_{6+ x }${P}$_{1- x }${M}$_{ x }${S}$_{5}${I} ({M} = {Si}, {Ge}, {Sn})},
  author={Ohno, Saneyuki and Helm, Bianca and Fuchs, Till and Dewald, Georg and Kraft, Marvin A and Culver, Sean P and Senyshyn, Anatoliy and Zeier, Wolfgang G},
  journal={Chemistry of Materials},
  volume={31},
  number={13},
  pages={4936--4944},
  year={2019},
  publisher={ACS Publications}
}

@article{31,
  title={Tailoring solution-processable {Li} argyrodites {Li}$_{6+ x }${P}$_{1- x }${M}$_{ x }${S}$_{5}${I} ({M} = {Ge}, {Sn}) and their microstructural evolution revealed by {Cryo}-{TEM} for all-solid-state batteries},
  author={Song, Yong Bae and Kim, Dong Hyeon and Kwak, Hiram and Han, Daseul and Kang, Sujin and Lee, Jong Hoon and Bak, Seong-Min and Nam, Kyung-Wan and Lee, Hyun-Wook and Jung, Yoon Seok},
  journal={Nano Letters},
  volume={20},
  number={6},
  pages={4337--4345},
  year={2020},
  publisher={ACS Publications}
}

@article{32,
  title={{Si}-doped {Li}$_{6}${PS}$_{5}${I} with enhanced conductivity enables superior performance for all-solid-state lithium batteries},
  author={Ming, Liang and Liu, Dan and Luo, Qiyue and Wei, Chaochao and Liu, Chen and Jiang, Ziling and Wu, Zhongkai and Li, Lin and Zhang, Long and Cheng, Shijie and others},
  journal={Chinese Chemical Letters},
  volume={35},
  number={10},
  pages={109387},
  year={2024},
  publisher={Elsevier}
}

@article{33,
  title={Enhanced air/electrochemical stability and Li-ion conductivity of argyrodite-type {Li}$_{5.5+ x }$({P}$_{1- x }${Ge}$_{ x }$){S}$_{4.5}${Cl}$_{1.5}$ ceramic solid electrolytes for all-solid-state batteries},
  author={Li, Haoze and Liu, Hao and Ding, Honggeng and Fu, Chenshuo and Gao, Ling and Li, Zhe and Liu, Haijing and Zhang, Yi and Khan, Abdul Jabbar and Zhao, Guowei},
  journal={Journal of Alloys and Compounds},
  volume={1026},
  pages={180413},
  year={2025},
  publisher={Elsevier}
}

@article{34,
  title={Lithium-site substituted argyrodite-type {Li}$_{6}${PS}$_{5}${I} solid electrolytes with enhanced ionic conduction for all-solid-state batteries},
  author={Gao, Ling and Xie, YuLin and Tong, Yan and Xu, Miao and You, JiaLe and Wei, HuiPing and Yu, XiangXiang and Xu, SiQi and Zhang, Yi and Che, Yong and others},
  journal={Science China Technological Sciences},
  volume={66},
  number={7},
  pages={2059--2068},
  year={2023},
  publisher={Springer}
}

@article{35,
  title={Substitutional disorder: structure and ion dynamics of the argyrodites {Li}$_{6}${PS}$_{5}${Cl}, {Li}$_{6}${PS}$_{5}${Br} and {Li}$_{6}${PS}$_{5}${I}},
  author={Hanghofer, I and Brinek, M and Eisbacher, SL and Bitschnau, B and Volck, M and Hennige, V and Hanzu, I and Rettenwander, D and Wilkening, HMR},
  journal={Physical Chemistry Chemical Physics},
  volume={21},
  number={16},
  pages={8489--8507},
  year={2019},
  publisher={Royal Society of Chemistry}
}

@article{36,
  title={Achieving superior ionic conductivity of {Li}$_{6}${PS}$_{5}${I} via introducing {LiCl}},
  author={Liao, Cong and Yu, Chuang and Peng, Linfeng and Miao, Xuefei and Chen, Shuai and Zhang, Ziqi and Cheng, Shijie and Xie, Jia},
  journal={Solid State Ionics},
  volume={377},
  pages={115871},
  year={2022},
  publisher={Elsevier}
}

@article{37,
  title={{Li}$_{6}${PS}$_{5}${X}: a class of crystalline {Li}-rich solids with an unusually high {Li}$^+$ mobility},
  author={Deiseroth, Hans-J{\"o}rg and Kong, Shiao-Tong and Eckert, Hellmut and Vannahme, Julia and Reiner, Christof and Zai{\ss}, Torsten and Schlosser, Marc},
  journal={Angewandte Chemie},
  volume={120},
  number={4},
  pages={767--770},
  year={2008},
  publisher={WILEY-VCH Verlag Weinheim}
}

@article{38,
  title={Structural Disorder in {Li}$_{6}${PS}$_{5}${I} Speeds {\(^{7}\)}{Li} Nuclear Spin Recovery and Slows Down {\(^{31}\)}{P} Relaxation--Implications for Translational and Rotational Jumps as Seen by Nuclear Magnetic Resonance},
  author={Brinek, M and Hiebl, C and Hogrefe, K and Hanghofer, I and Wilkening, HMR},
  journal={The Journal of Physical Chemistry C},
  volume={124},
  number={42},
  pages={22934--22940},
  year={2020},
  publisher={ACS Publications}
}

@article{39,
  title={New family of argyrodite thioantimonate lithium superionic conductors},
  author={Zhou, Laidong and Assoud, Abdeljalil and Zhang, Qiang and Wu, Xiaohan and Nazar, Linda F},
  journal={Journal of the American Chemical Society},
  volume={141},
  number={48},
  pages={19002--19013},
  year={2019},
  publisher={ACS Publications}
}

@article{40,
  title={Microscopic degradation mechanism of argyrodite-type sulfide at the solid electrolyte--cathode interface},
  author={Morino, Yusuke and Tsukasaki, Hirofumi and Mori, Shigeo},
  journal={ACS Applied Materials \& Interfaces},
  volume={15},
  number={19},
  pages={23051--23057},
  year={2023},
  publisher={ACS Publications}
}

@article{41,
  title={Constructing {Br}-doped {Li}$_{10}${SnP}$_{2}${S}$_{12}$-based all-solid-state batteries with superior performances},
  author={Luo, Qiyue and Ming, Liang and Zhang, Dong and Wei, Chaochao and Wu, Zhongkai and Jiang, Ziling and Liu, Chen and Liu, Shiyu and Cao, Kecheng and Zhang, Long and others},
  journal={Energy Material Advances},
  volume={4},
  pages={0065},
  year={2023},
  publisher={AAAS}
}

@article{42,
  title={Unravelling the {O}-doping effect on chemical/electrochemical stability of {Li}$_{5.5}${PS}$_{4.5}${Cl}$_{1.5}$ for all-solid-state lithium batteries},
  author={Ming, Liang and Luo, Qiyue and Wei, Chaochao and Liu, Chen and Jiang, Ziling and Wu, Zhongkai and Li, Lin and Zhang, Long and Chen, Xia and Cheng, Shijie and others},
  journal={Next Materials},
  volume={5},
  pages={100233},
  year={2024},
  publisher={Elsevier}
}

@article{43,
  title={Reviving the ionic conductivity of air-instable solid-state electrolytes via a facile heat treatment},
  author={Ming, Liang and Deng, Miao and Li, Siwu and Jiang, Ziling and Li, Lin and Lu, Ziyu and Luo, Qiyue and Yang, Jie and Cui, Zhonghui and Yu, Chuang},
  journal={Chinese Chemical Letters},
  pages={111114},
  year={2025},
  publisher={Elsevier}
}

@article{44,
  title={Lithium-ion battery: A comprehensive research progress of high nickel ternary cathode material},
  author={Chang, Longjiao and Wei, Anlu and Luo, Shaohua and Cao, Shiyuan and Bi, Xiaolong and Yang, Wei and Yang, Ruifen and Liu, Jianan},
  journal={International Journal of Energy Research},
  volume={46},
  number={15},
  pages={23145--23172},
  year={2022},
  publisher={Wiley Online Library}
}

@article{45,
  title={All-solid-state lithium batteries with sulfide electrolytes and oxide cathodes},
  author={Wu, Jinghua and Shen, Lin and Zhang, Zhihua and Liu, Gaozhan and Wang, Zhiyan and Zhou, Dong and Wan, Hongli and Xu, Xiaoxiong and Yao, Xiayin},
  journal={Electrochemical Energy Reviews},
  volume={4},
  number={1},
  pages={101--135},
  year={2021},
  publisher={Springer}
}

@article{46,
  title={Electrochemical and material analyses for sulfide-based solid electrolyte--cathode interface under high voltage},
  author={Morino, Yusuke and Kanada, Satoshi},
  journal={Journal of Power Sources},
  volume={509},
  pages={230376},
  year={2021},
  publisher={Elsevier}
}

@article{47,
  title={Charged and discharged states of cathode/sulfide electrolyte interfaces in all-solid-state lithium ion batteries},
  author={Sumita, Masato and Tanaka, Yoshinori and Ikeda, Minoru and Ohno, Takahisa},
  journal={The Journal of Physical Chemistry C},
  volume={120},
  number={25},
  pages={13332--13339},
  year={2016},
  publisher={ACS Publications}
}

@article{48,
  title={Outstanding electrochemical performances of the all-solid-state lithium battery using {Ni}-rich layered oxide cathode and sulfide electrolyte},
  author={Li, Xuelei and Sun, Qifang and Wang, Zhengyu and Song, Dawei and Zhang, Hongzhou and Shi, Xixi and Li, Chunliang and Zhang, Lianqi and Zhu, Lingyun},
  journal={Journal of Power Sources},
  volume={456},
  pages={227997},
  year={2020},
  publisher={Elsevier}
}

@article{49,
  title={Space--charge layer effect at interface between oxide cathode and sulfide electrolyte in all-solid-state lithium-ion battery},
  author={Haruyama, Jun and Sodeyama, Keitaro and Han, Liyuan and Takada, Kazunori and Tateyama, Yoshitaka},
  journal={Chemistry of Materials},
  volume={26},
  number={14},
  pages={4248--4255},
  year={2014},
  publisher={ACS Publications}
}

@article{50,
  title={Interface design considering intrinsic properties of dielectric materials to minimize space-charge layer effect between oxide cathode and sulfide solid electrolyte in all-solid-state batteries},
  author={Park, Bo Keun and Kim, Hyeongil and Kim, Kyung Su and Kim, Hyun-Seung and Han, Seung Ho and Yu, Ji-Sang and Hah, Hoe Jin and Moon, Janghyuk and Cho, Woosuk and Kim, Ki Jae},
  journal={Advanced Energy Materials},
  volume={12},
  number={37},
  pages={2201208},
  year={2022},
  publisher={Wiley Online Library}
}

@article{51,
  title={Elucidating and minimizing the space-charge layer effect between NCM cathode and {Li}$_{6}${PS}$_{5}${Cl} for sulfide-based solid-state lithium batteries},
  author={Chen, Ya and Huang, Ling and Zhou, Deli and Gao, Xin and Hu, Tengfei and Zhang, Zhiyuan and Zhen, Zheng and Chen, Xiaodong and Cui, Lifeng and Wang, Guoxiu},
  journal={Advanced Energy Materials},
  volume={14},
  number={30},
  pages={2304443},
  year={2024},
  publisher={Wiley Online Library}
}

@article{52,
  title={In-situ visualization of the space-charge-layer effect on interfacial lithium-ion transport in all-solid-state batteries},
  author={Wang, Longlong and Xie, Ruicong and Chen, Bingbing and Yu, Xinrun and Ma, Jun and Li, Chao and Hu, Zhiwei and Sun, Xingwei and Xu, Chengjun and Dong, Shanmu and others},
  journal={Nature Communications},
  volume={11},
  number={1},
  pages={5889},
  year={2020},
  publisher={Nature Publishing Group UK London}
}

@Article{53,
author={Zhao, Zirui and Li, Hai-Feng},
title="{Investigating Material Interface Diffusion Phenomena through Graph Neural Networks in Applied Materials}",
journal={ACS Applied Materials {\&} Interfaces},
year={2024},
publisher={American Chemical Society},
volume={16},
number={39},
pages={53153-53162}
}

@article{54,
  title={Electrochemical properties of cathode according to the type of sulfide electrolyte and the application of surface coating},
  author={Yoon, Da Hye and Park, Yong Joon},
  journal={Journal of Electrochemical Science and Technology},
  volume={12},
  number={1},
  pages={126--136},
  year={2020},
  publisher={The Korean Electrochemical Society}
}

@article{55,
  title={{Multifunctional Coatings on Sulfide-Based Solid Electrolyte Powders with Enhanced Processability, Stability, and Performance for Solid-State Batteries}},
  author={Hood, Zachary D and Mane, Anil U and Sundar, Aditya and Tepavcevic, Sanja and Zapol, Peter and Eze, Udochukwu D and Adhikari, Shiba P and Lee, Eungje and Sterbinsky, George E and Elam, Jeffrey W and others},
  journal={Advanced Materials},
  volume={35},
  number={21},
  pages={2300673},
  year={2023},
  publisher={Wiley Online Library}
}

@article{56,
  title={Stabilized cathode/sulfide electrolyte interface through conformally interfacial nanocoating for all-solid-state batteries},
  author={Zou, Changfei and Zang, Zihao and Tao, Xiyuan and Yi, Lingguang and Chen, Xiaoyi and Zhang, Xiaoyan and Yang, Li and Liu, Xianhu and Wang, Xianyou},
  journal={ACS Applied Energy Materials},
  volume={6},
  number={6},
  pages={3599--3607},
  year={2023},
  publisher={ACS Publications}
}

@article{57,
  title={{O}-doping strategy enabling enhanced chemical/electrochemical stability of {Li}$_{3}${InCl}$_{6}$ for superior solid-state battery performance},
  author={Luo, Qiyue and Liu, Chen and Li, Lin and Jiang, Ziling and Yang, Jie and Chen, Shaoqing and Chen, Xia and Zhang, Long and Cheng, Shijie and Yu, Chuang},
  journal={Journal of Energy Chemistry},
  volume={99},
  pages={484--494},
  year={2024},
  publisher={Elsevier}
}

@article{58,
  title={{Local Charge Distribution Regulation toward Sulfide Superionic Conductor for Superior Electrochemical Performance in All-Solid-State Batteries}},
  author={Luo, Qiyue and Li, Siwu and Wei, Chaochao and Wu, Zhongkai and Liu, Chen and Jiang, Ziling and Li, Lin and Chen, Shaoqing and Yu, Chuang},
  journal={ACS Applied Energy Materials},
  volume={8},
  number={4},
  pages={2465--2476},
  year={2025},
  publisher={ACS Publications}
}

@article{59,
  title={Visualizing the chemical incompatibility of halide and sulfide-based electrolytes in solid-state batteries},
  author={Rosenbach, Carolin and Walther, Felix and Ruhl, Justine and Hartmann, Matthias and Hendriks, Theodoor Anton and Ohno, Saneyuki and Janek, J{\"u}rgen and Zeier, Wolfgang G},
  journal={Advanced Energy Materials},
  volume={13},
  number={6},
  pages={2203673},
  year={2023},
  publisher={Wiley Online Library}
}

@article{60,
  title={Halide/sulfide composite solid-state electrolyte for {Li}-anode based all-solid-state batteries},
  author={Zhang, Haochang and Yu, Zhaozhe and Cheng, Jinyin and Chen, Hannan and Huang, Xiao and Tian, Bingbing},
  journal={Chinese Chemical Letters},
  volume={34},
  number={11},
  pages={108228},
  year={2023},
  publisher={Elsevier}
}

@article{61,
  title={Control of side reactions using {LiNbO}$_{3}$ mixed/doped solid electrolyte for enhanced sulfide-based all-solid-state batteries},
  author={Cho, Ji-Un and Rajagopal, Rajesh and Yoon, Da Hye and Park, Yong Joon and Ryu, Kwang-Sun},
  journal={Chemical Engineering Journal},
  volume={452},
  pages={138955},
  year={2023},
  publisher={Elsevier}
}

@article{62,
  title={Enhanced cathode/sulfide electrolyte interface stability using an {Li}$_{2}${ZrO}$_{3}$ coating for all-solid-state batteries},
  author={Lee, Jun Won and Park, Yong Joon},
  journal={Journal of Electrochemical Science and Technology},
  volume={9},
  number={3},
  pages={176--183},
  year={2018},
  publisher={The Korean Electrochemical Society}
}

@article{63,
  title={Novel dry deposition of {LiNbO}$_{3}$ or {Li}$_{2}${ZrO}$_{3}$ on {LiNi}$_{0.6}${Co}$_{0.2}${Mn}$_{0.2}${O}$_{2}$ for high performance all-solid-state lithium batteries},
  author={Kim, Young-Jin and Rajagopal, Rajesh and Kang, Sung and Ryu, Kwang-Sun},
  journal={Chemical Engineering Journal},
  volume={386},
  pages={123975},
  year={2020},
  publisher={Elsevier}
}

@article{64,
  title={{Oxygen-Substitution Effects on the Properties of Argyrodite-Type Sulfide Solid Electrolytes ({Li}$_{5.5}${PS}$_{4.5- x}${Br}$_{1.5}${O}$_{x}$, $0 \leq  x \leq  0.5$)}},
  author={Tsukazaki, Rei and Matsui, Naoki and Hori, Satoshi and Suzuki, Kota and Kanno, Ryoji},
  journal={Electrochemistry},
  volume={93},
  number={6},
  pages={063012--063012},
  year={2025},
  publisher={The Electrochemical Society of Japan}
}

@article{65,
  title={Recrystallization and heterovalent substitution effects on mechanical and electrical parameters of {Ag}$_{6+x}$({P}$_{1-x}${Ge}$_{ x }$){S}$_{5}${I}--based ceramics},
  author={Pogodin, Artem and Filep, Mykhailo and Malakhovska, Tetyana and Vakulchak, Vasyl and Komanicky, Vladimir and Vorobiov, Serhii and Izai, Vitalii and Satrapinskyy, Leonid and Shender, Iryna and Bilanych, Vitaliy and others},
  journal={Journal of the European Ceramic Society},
  volume={44},
  number={6},
  pages={4097--4110},
  year={2024},
  publisher={Elsevier}
}

@article{66,
  title={Quantitative analysis of crystallinity in an argyrodite sulfide-based solid electrolyte synthesized via solution processing},
  author={Yubuchi, So and Tsukasaki, Hirofumi and Sakuda, Atsushi and Mori, Shigeo and Hayashi, Akitoshi and Tatsumisago, Masahiro},
  journal={RSC Advances},
  volume={9},
  number={25},
  pages={14465--14471},
  year={2019},
  publisher={Royal Society of Chemistry}
}

@article{67,
  title={Enhancing ionic conductivity by in situ formation of {Li}$_{7}${SiPS}$_{8}$/{Argyrodite} hybrid solid electrolytes},
  author={Calaminus, Robert and Harm, Sascha and Fabini, Douglas H and Balzat, Lucas G and Hatz, Anna-Katharina and Duppel, Viola and Moudrakovski, Igor and Lotsch, Bettina V},
  journal={Chemistry of Materials},
  volume={34},
  number={17},
  pages={7666--7677},
  year={2022},
  publisher={ACS Publications}
}

@article{68,
  title={Mitigation of the instability of ultrafast {Li}-ion conductor {Li}$_{6.6}${Si}$_{0.6}${Sb}$_{0.4}${S}$_{5}${I} enables high-performance all-solid-state batteries},
  author={Liao, Cong and Yu, Chuang and Chen, Shaoqing and Wei, Chaochao and Wu, Zhongkai and Chen, Shuai and Jiang, Ziling and Cheng, Shijie and Xie, Jia},
  journal={Renewables},
  volume={1},
  number={2},
  pages={266--276},
  year={2023},
  publisher={Chinese Chemical Society Zhongguancun, Haidian, Beijing 100190, China}
}

@article{69,
  title={Unraveling the {Li}{NbO}$_{3}$ coating layer on battery performances of lithium argyrodite-based all-solid-state batteries under different cut-off voltages},
  author={Wei, Chaochao and Yu, Chuang and Chen, Shaoqing and Chen, Shuai and Peng, Linfeng and Wu, Yuanke and Li, Shuping and Cheng, Shijie and Xie, Jia},
  journal={Electrochimica Acta},
  volume={438},
  pages={141545},
  year={2023},
  publisher={Elsevier}
}

@article{70,
  title={Effect of amorphous LiPON coating on electrochemical performance of {LiNi}$_{0.8}${Mn}$_{0.1}${Co}$_{0.1}${O}$_{2}$ ({NMC811}) in all solid-state batteries},
  author={Shrestha, Sushovan and Kim, Jongbeon and Jeong, Jejun and Lee, Hye Jin and Kim, Seul Cham and Hah, Hoe Jin and Oh, Kyuhwan and Lee, Se-Hee},
  journal={Journal of The Electrochemical Society},
  volume={168},
  number={6},
  pages={060537},
  year={2021},
  publisher={IOP Publishing}
}

@article{71,
  title={{Surface-to-Bulk Synergistic Modification of Single Crystal Cathode Enables Stable Cycling of Sulfide-Based All-Solid-State Batteries at 4.4 V}},
  author={Sun, Nan and Song, Yajie and Liu, Qingsong and Zhao, Wei and Zhang, Fang and Ren, Liping and Chen, Ming and Zhou, Zinan and Xu, Zihan and Lou, Shuaifeng and others},
  journal={Advanced Energy Materials},
  volume={12},
  number={29},
  pages={2200682},
  year={2022},
  publisher={Wiley Online Library}
}

@article{72,
  title={{An Anthraquinone/Carbon Fiber Composite as Cathode Material for Rechargeable Sodium-Ion Batteries}},
  author={Werner, Daniel and Apaydin, Dogukan H and Portenkirchner, Engelbert},
  journal={Batteries \& Supercaps},
  volume={1},
  number={4},
  pages={160--168},
  year={2018},
  publisher={Wiley Online Library}
}

@article{73,
  title={{Electrochemical Study on Rate Characteristics of All-Solid-State Batteries}},
  author={Joo, Soyoung and Kim, Seong-Yoon and Min, Yu-Jeong and Kim, Seong-Yoon},
  booktitle={Electrochemical Society Meeting Abstracts prime2024},
  number={10},
  pages={4963--4963},
  year={2024},
  organization={The Electrochemical Society, Inc.}
}

@article{74,
  title={Understanding the carbon additive/sulfide solid electrolyte interface in nickel-rich cathode composites and prioritizing the corresponding interplay between the electrical and ionic conductive networks to enhance all-solid-state-battery rate capability},
  author={Saqib, Kashif Saleem and Embleton, Tom James and Choi, Jae Hong and Won, Sung-Jae and Ali, Jahanzaib and Ko, Kyungmok and Choi, Sumyeong and Jo, Mina and Park, Sungwoo and Park, Joohyuk and others},
  journal={ACS Applied Materials \& Interfaces},
  volume={16},
  number={36},
  pages={47551--47562},
  year={2024},
  publisher={ACS Publications}
}

@article{75,
  title={Enabling Argyrodite Sulfides as Superb Solid-State Electrolyte with Remarkable Interfacial Stability Against Electrodes},
  author={Xu, Hongjie and Cao, Guoqin and Shen, Yonglong and Yu, Yuran and Hu, Junhua and Wang, Zhuo and Shao, Guosheng},
  journal={Energy \& Environmental Materials},
  volume={5},
  number={3},
  pages={852--864},
  year={2022},
  publisher={Wiley Online Library}
}

@article{76,
  title={Degradation analysis during fast lifetime cycling of sulfide-based all-solid-state {Li}-metal batteries using in situ electrochemical impedance spectroscopy},
  author={Kim, Young Jung and Jeong, Hyeseong and Nam, Sahn and Shin, Dongwook and Lee, Jong-Ho and Kim, Hyoungchul},
  journal={Journal of Materials Chemistry A},
  year={2025},
  publisher={Royal Society of Chemistry}
}

@article{77,
  title={Conductivity Enhancement of Argyrodite {Li}$_6${SbS}$_5${I} Solid Electrolyte via Charge Modulation Around {Li} Diffusion Paths Through {Si} Substitution},
  author={Yi, Seho and Jeon, Taegon and Lee, Jaeho and Han, Young-Kyu and Jung, Sung Chul},
  journal={Journal of Materials Chemistry A},
  volume={13},
  pages={36597--36608},
  year={2025},
  publisher={Royal Society of Chemistry}
}

\appendix
\clearpage
\section*{Appendix}

\clearpage 

\renewcommand{\thefigure}{S\arabic{figure}}
\setcounter{figure}{0} 

\begin{figure*}
\centering
\includegraphics[width=0.92\textwidth]{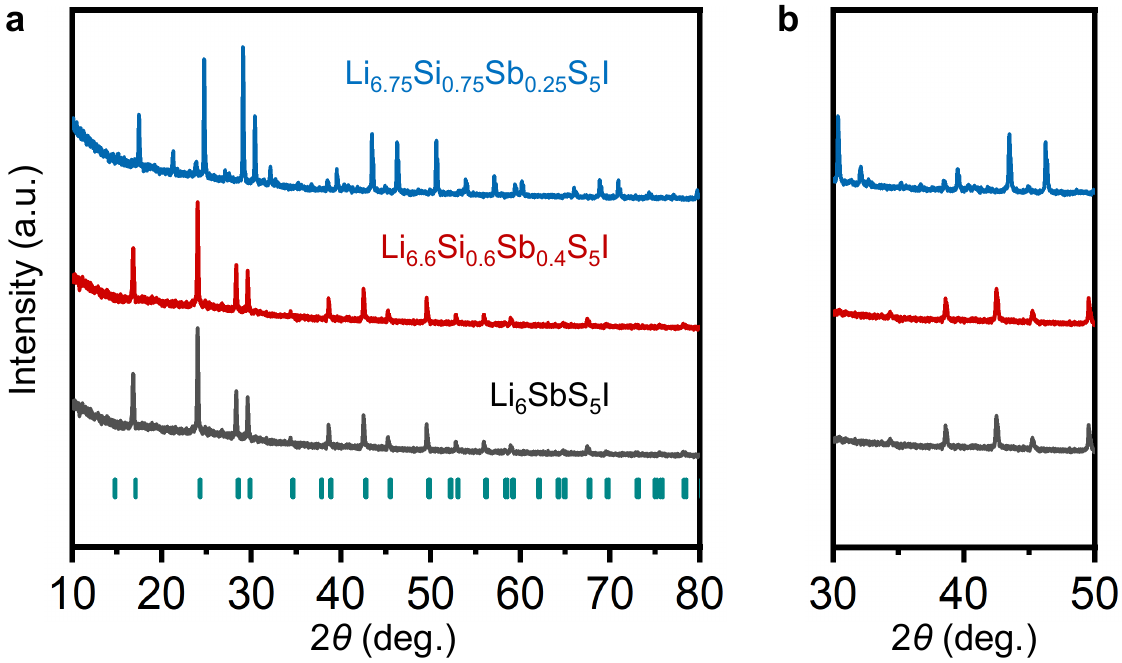}
\caption{Structural evolution with varying Si content. 
(a) XRD spectra of Li$_{6+x}$Si$_{x}$Sb$_{1-x}$S$_5$I ($x = 0$, 0.6, 0.75), and (b) the magnified patterns within 30--50$^\circ$, revealing distinct structural features and phase evolution with increasing Si content.}
\label{Fig. S1}
\end{figure*}

\clearpage

\begin{figure*}
\centering
\includegraphics[width=0.92\textwidth]{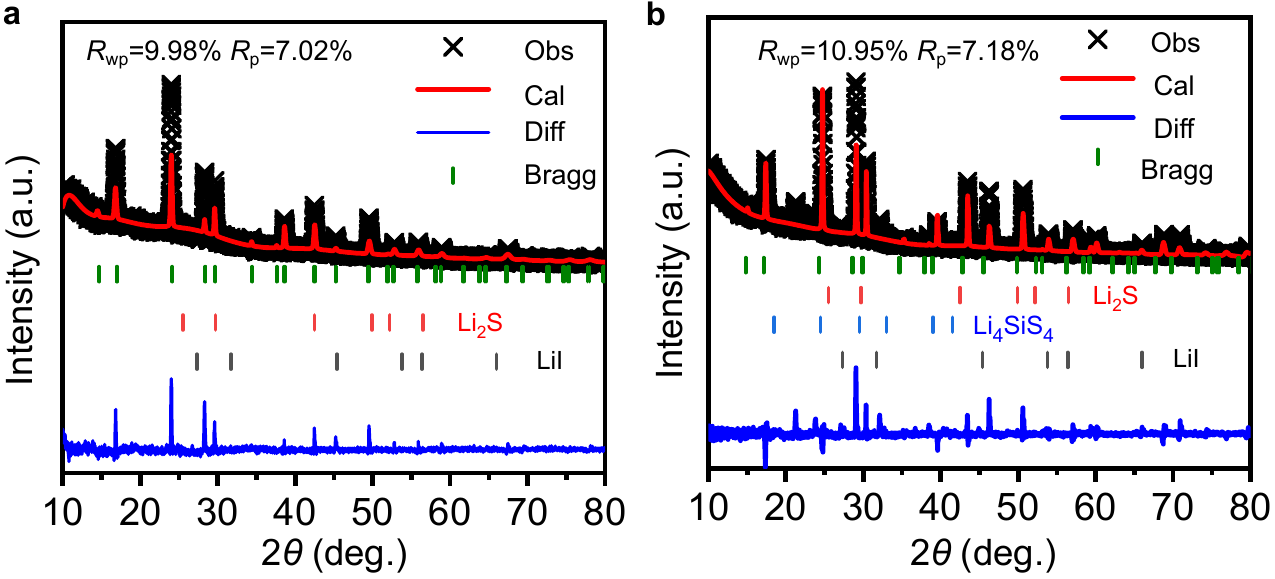}
\caption{Rietveld refinement of electrolyte structures. 
XRD Rietveld refinement patterns for (a) Li$_6$SbS$_5$I and (b) Li$_{6.75}$Si$_{0.75}$Sb$_{0.25}$S$_5$I.}
\label{Fig. S2}
\end{figure*}

\clearpage

\begin{figure*}
\centering
\includegraphics[width=0.92\textwidth]{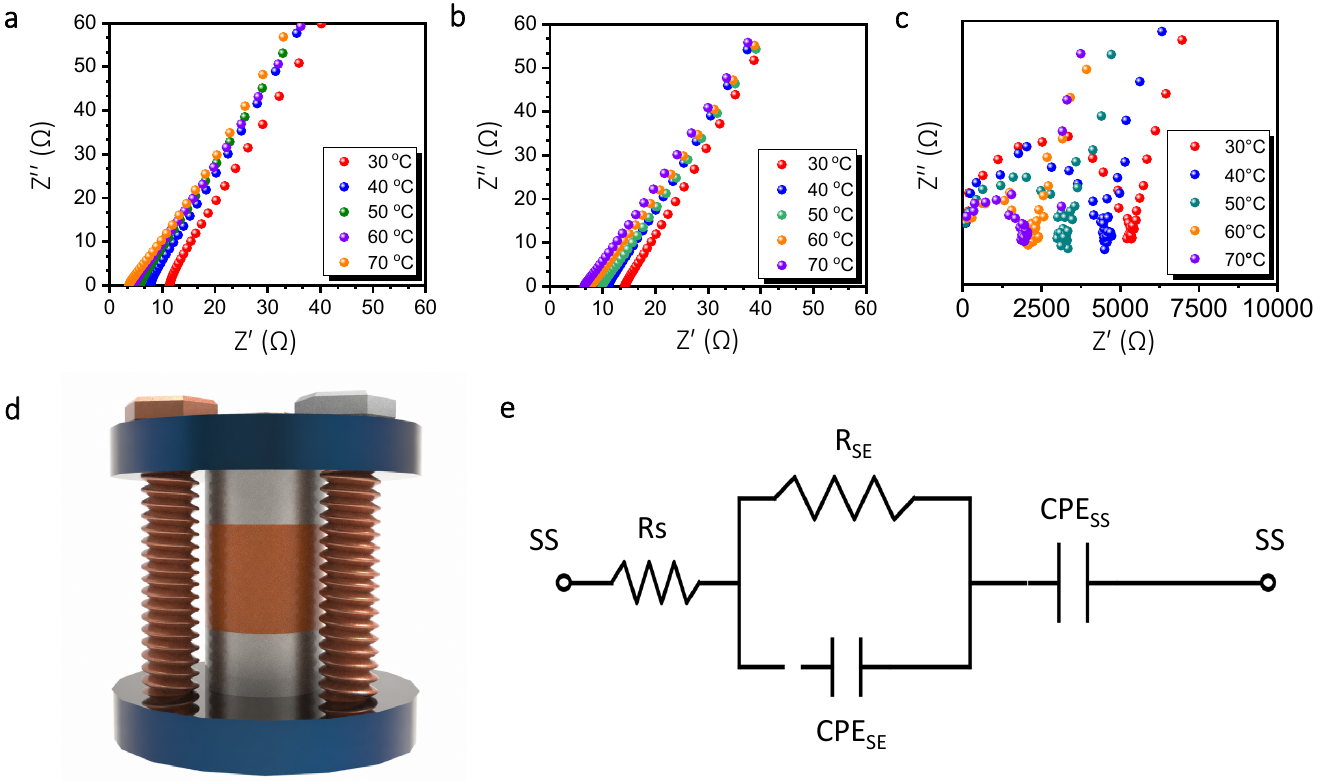}
\caption{Temperature-dependent EIS spectra of Li$_{6+x}$Si$_x$Sb$_{1-x}$S$_5$I electrolytes measured from 30 to 70~$^\circ$C: (a) Li$_{6.6}$Si$_{0.6}$Sb$_{0.4}$S$_5$I, (b) Li$_{6.75}$Si$_{0.75}$Sb$_{0.25}$S$_5$I, and (c) Li$_6$SbS$_5$I. (d) Schematic illustration of the SS $\vert$ electrolyte $\vert$ SS blocking-cell configuration used for EIS measurements with (e) the corresponding equivalent-circuit model.}
\label{Fig. S3}
\end{figure*}

\clearpage

\begin{figure*}
\centering
\includegraphics[width=0.92\textwidth]{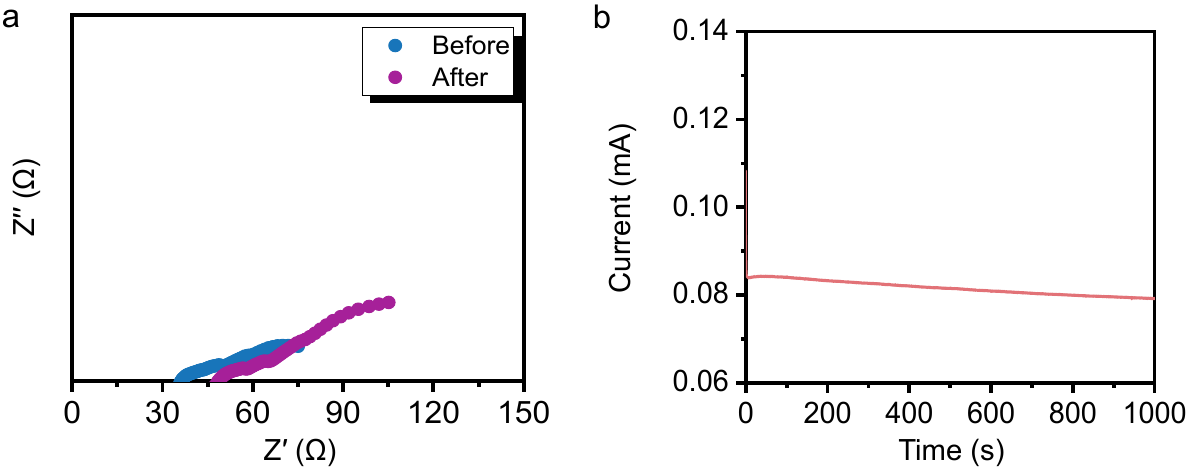}
\caption{Lithium-ion transference number measurement of the optimized Li$_{6.6}$Si$_{0.6}$Sb$_{0.4}$S$_5$I electrolyte: (a) electrochemical impedance spectra collected before and after DC polarization for the Li | Li$_{6.6}$Si$_{0.6}$Sb$_{0.4}$S$_5$I | Li symmetric cell; (b) current--time curve recorded during chronoamperometric polarization.}
\label{Fig. S4}
\end{figure*}

\clearpage

\begin{figure*}
\centering
\includegraphics[width=0.92\textwidth]{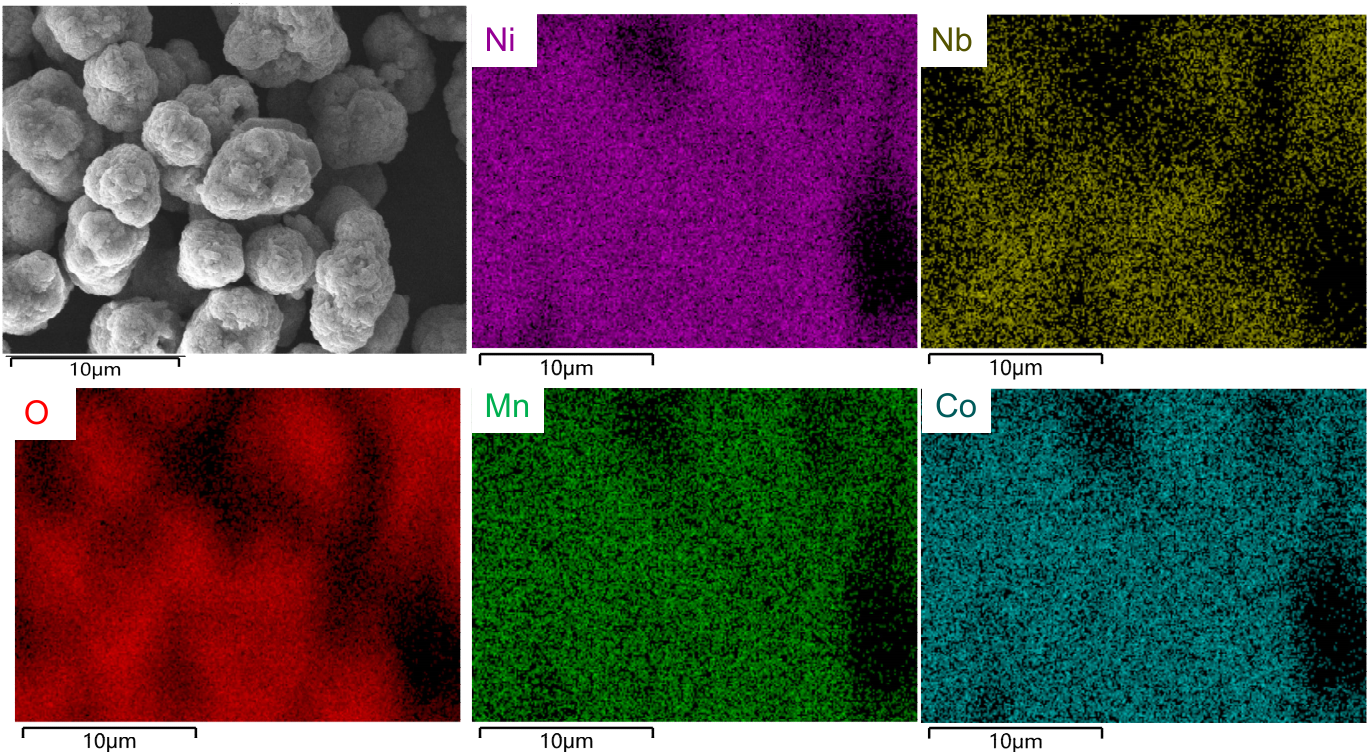}
\caption{SEM image and corresponding elemental mapping of LiNbO$_3$-coated NCM712 secondary particles.}
\label{Fig. S5}
\end{figure*}

\clearpage

\begin{figure*}
\centering
\includegraphics[width=0.92\textwidth]{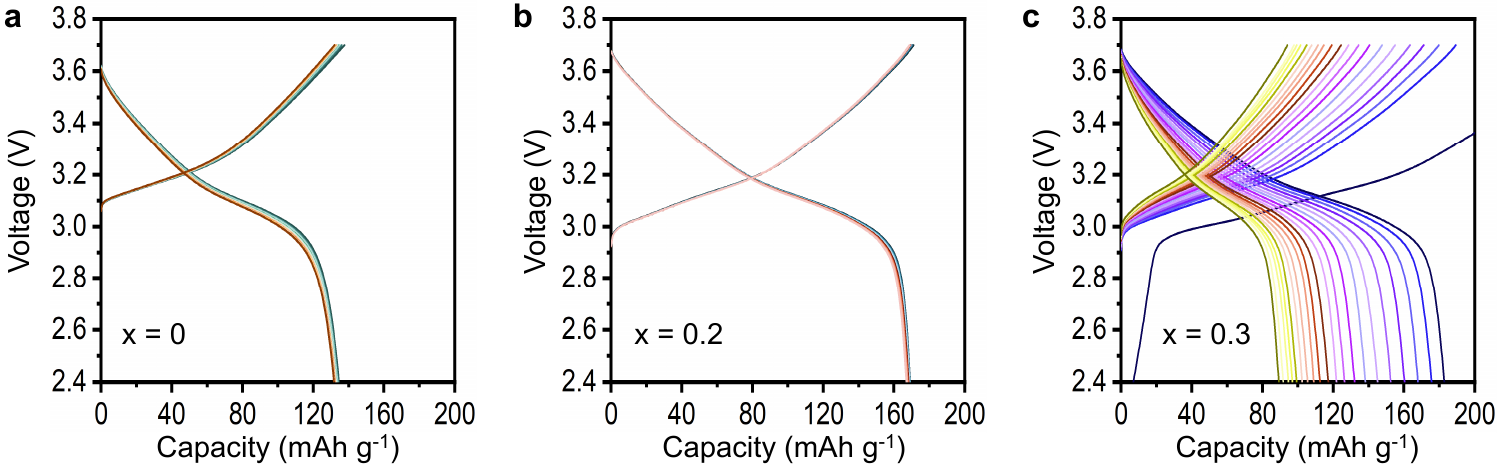}
\caption{Impedance analysis of composite cathodes. 
Impedance spectra of ASSBs employing composite cathodes with carbon ratios of (a) $x = 0$, (b) $x = 0.2$, and (c) $x = 0.3$, highlighting the impact of carbon content on interfacial resistance and ion transport.}
\label{Fig. S6}
\end{figure*}

\clearpage

\begin{figure*}
\centering
\includegraphics[width=0.92\textwidth]{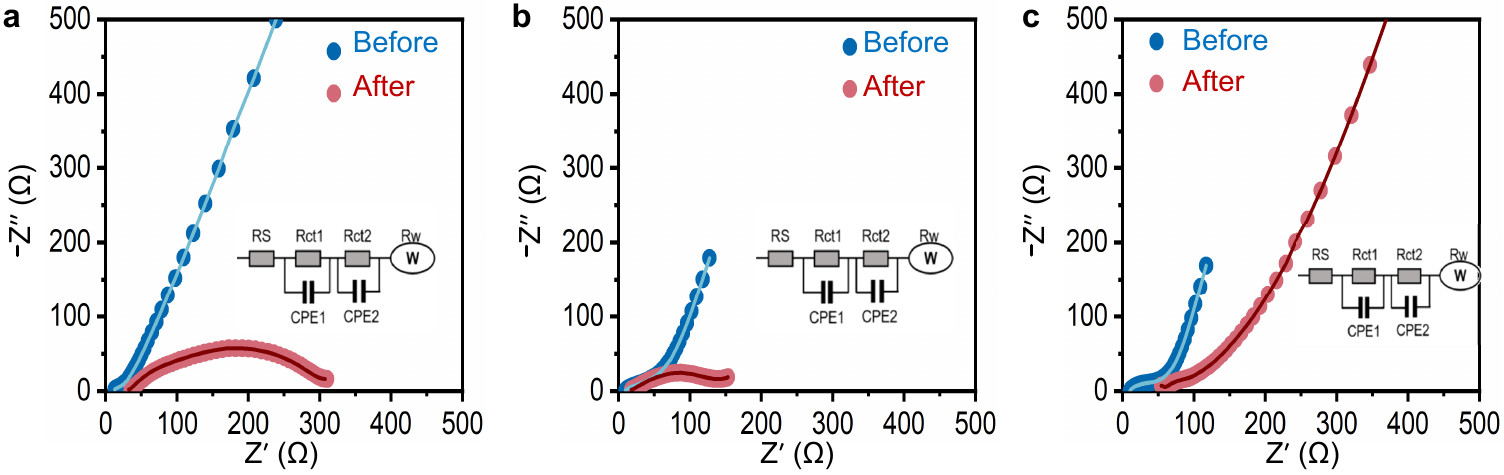}
\caption{Initial cycling behavior of composite cathodes. 
Charge/discharge curves during the initial 20 cycles of batteries utilizing composite cathodes with carbon ratios of (a) $x = 0$, (b) $x = 0.2$, and (c) $x = 0.3$, illustrating the influence of carbon addition on electrochemical performance and capacity retention.}
\label{Fig. S7}
\end{figure*}

\clearpage

\begin{figure*}
\centering
\includegraphics[width=0.92\textwidth]{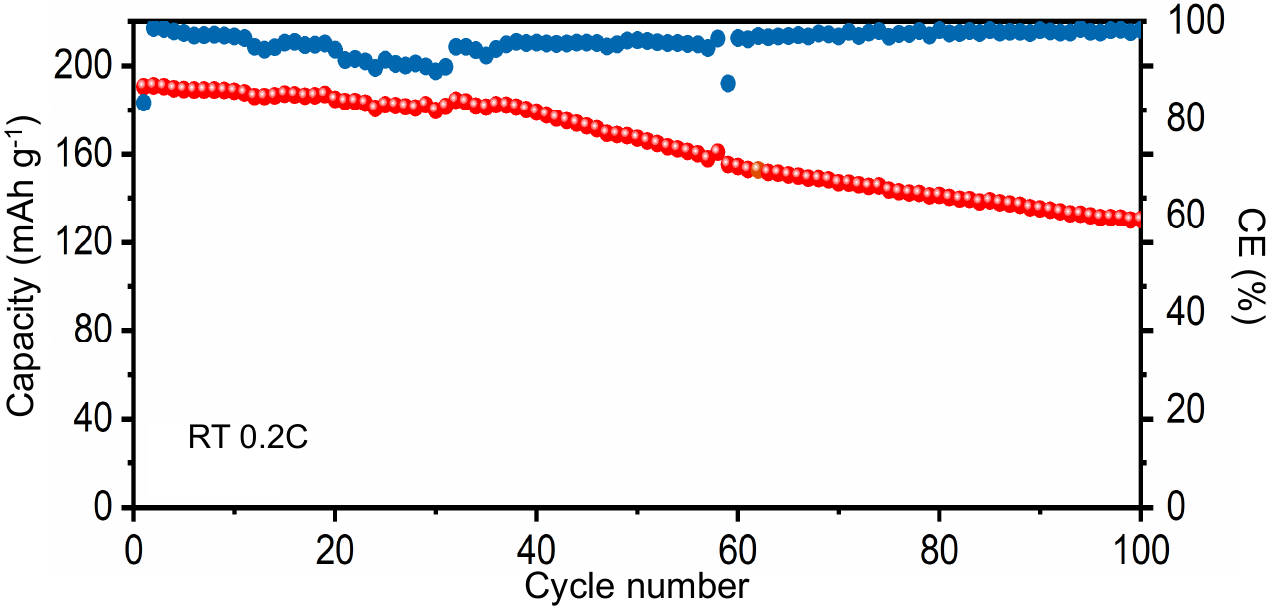}
\caption{Long-term stability of optimized cathodes. 
Long-term cycling performance of LNO@NCM-VGCF ($x = 0.2$)/Li$_{6.6}$Si$_{0.6}$Sb$_{0.4}$S$_5$I/Li-In at 0.2C, demonstrating outstanding stability and capacity retention over extended cycles.}
\label{Fig. S8}
\end{figure*}

\clearpage

\begin{figure*}
\centering
\includegraphics[width=0.92\textwidth]{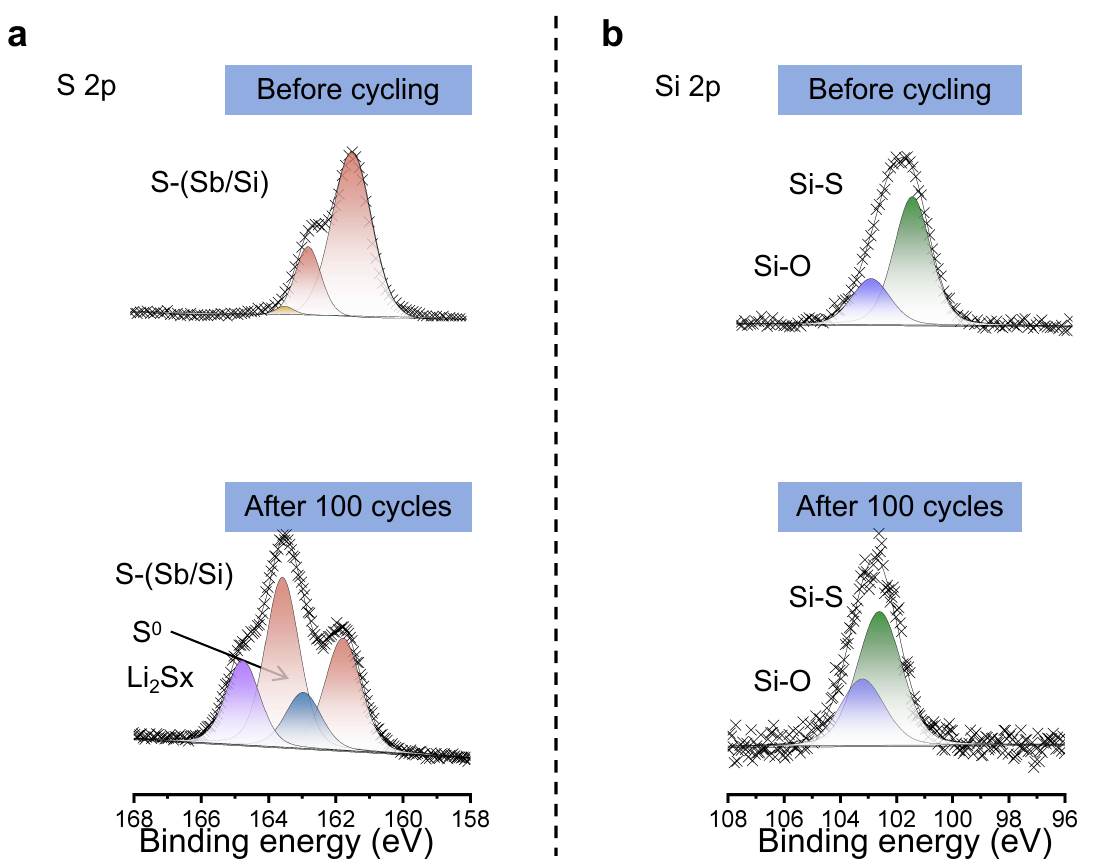}
\caption{XPS spectra of (a) S 2p and (b) Si 2p collected from pristine and cycled LNO@NCM712/Li$_{6.6}$Si$_{0.6}$Sb$_{0.4}$S$_5$I/Li--In cells with 3 wt\% VGCF in the composite cathode.}
\label{Fig. S9}
\end{figure*}

\clearpage

\begin{figure*}
\centering
\includegraphics[width=0.92\textwidth]{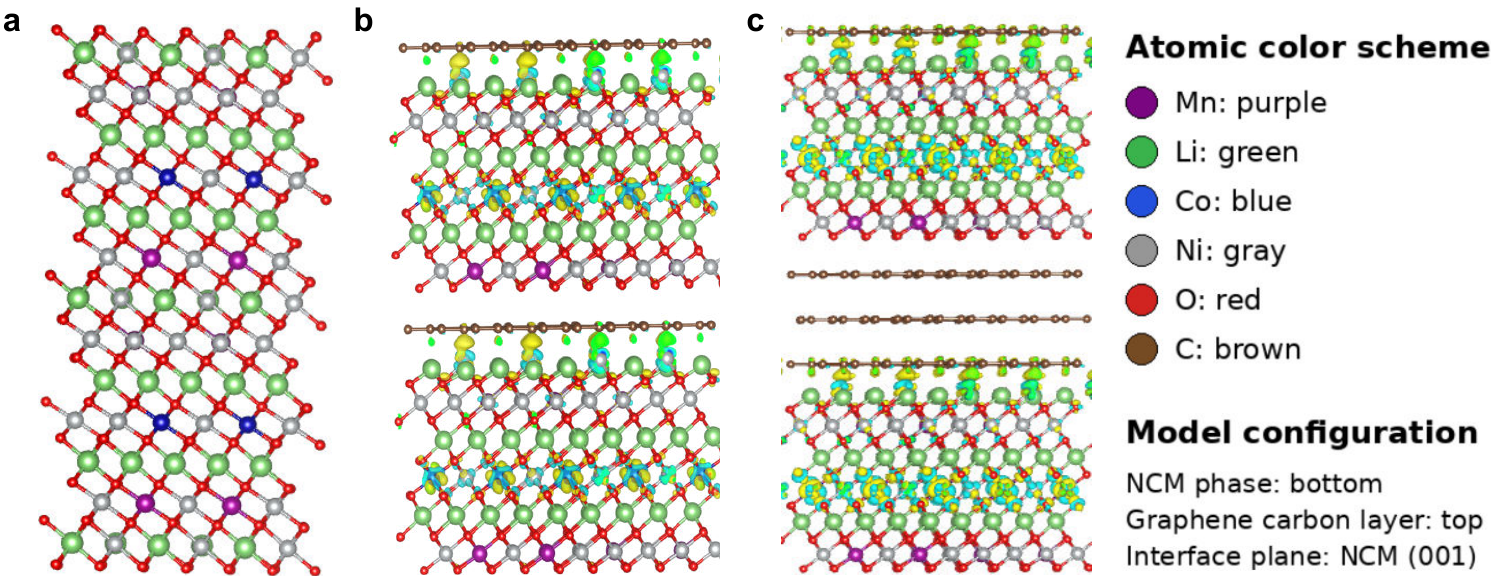}
\caption{Schematic diagram and charge density difference for LNO@NCM712/Li$_{6.6}$Si$_{0.6}$Sb$_{0.4}$S$_5$I/VGCF cathodes with VGCF amounts of (a) $x = 0$, (b) $x = 0.2$, and (c) $x = 0.3$.}
\label{Fig. S10}
\end{figure*}

\clearpage

\begin{figure*}
\centering
\includegraphics[width=0.92\textwidth]{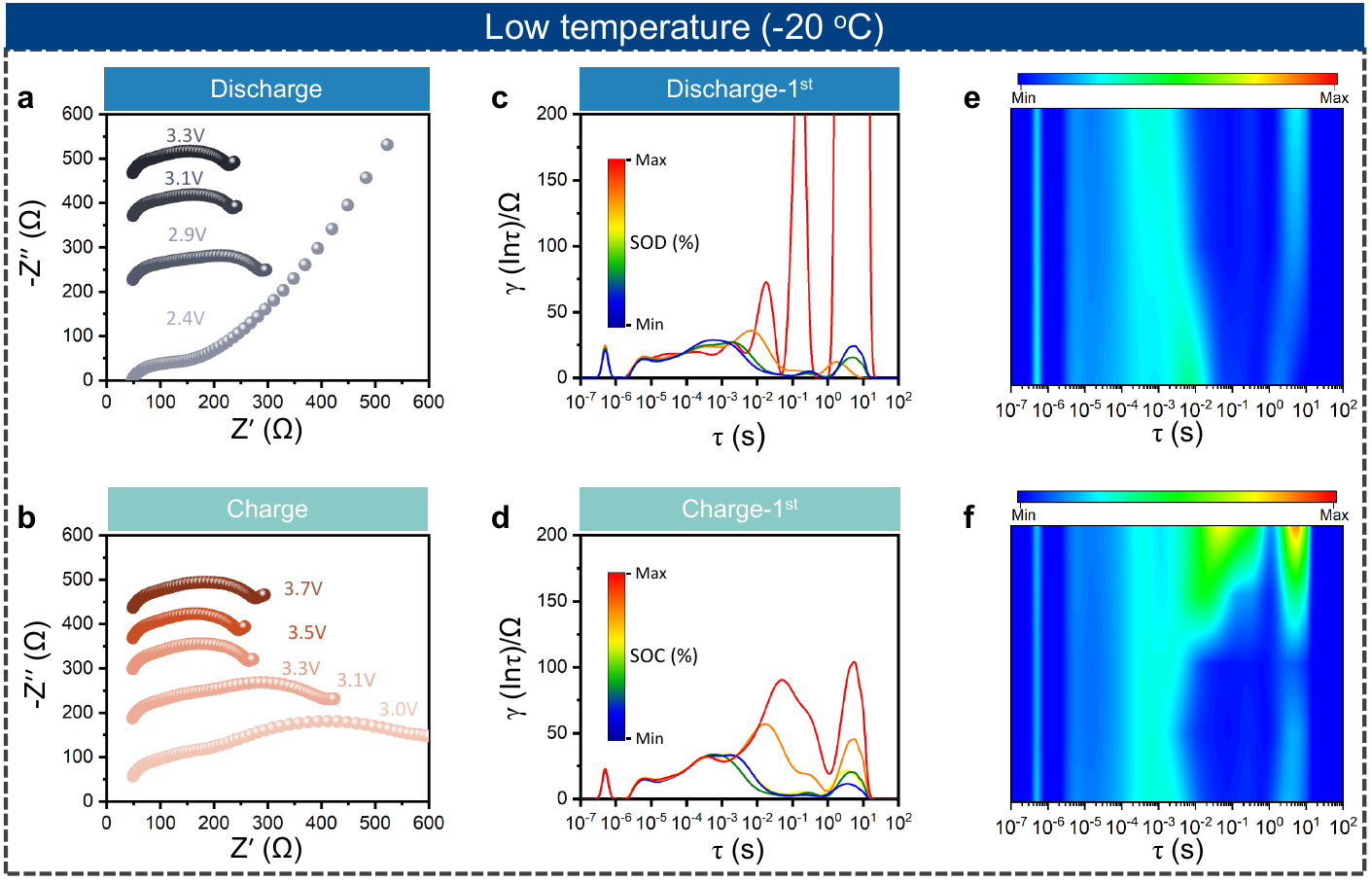}
\caption{Low-temperature interfacial dynamics. 
Electrochemical characterization of LNO@NCM712/Li$_{6.6}$Si$_{0.6}$Sb$_{0.4}$S$_5$I/Li-In batteries at a low temperature of $-20~^\circ$C during the initial cycle. Impedance spectra at various (a,b) charge (SOC) and discharge (SOD) states at low temperature, with corresponding (c,d) DRT curves and (e,f) 2D intensity maps, revealing interfacial processes and temperature-dependent dynamics.}
\label{Fig. S11}
\end{figure*}

\clearpage

\begin{figure*}
\centering
\includegraphics[width=0.92\textwidth]{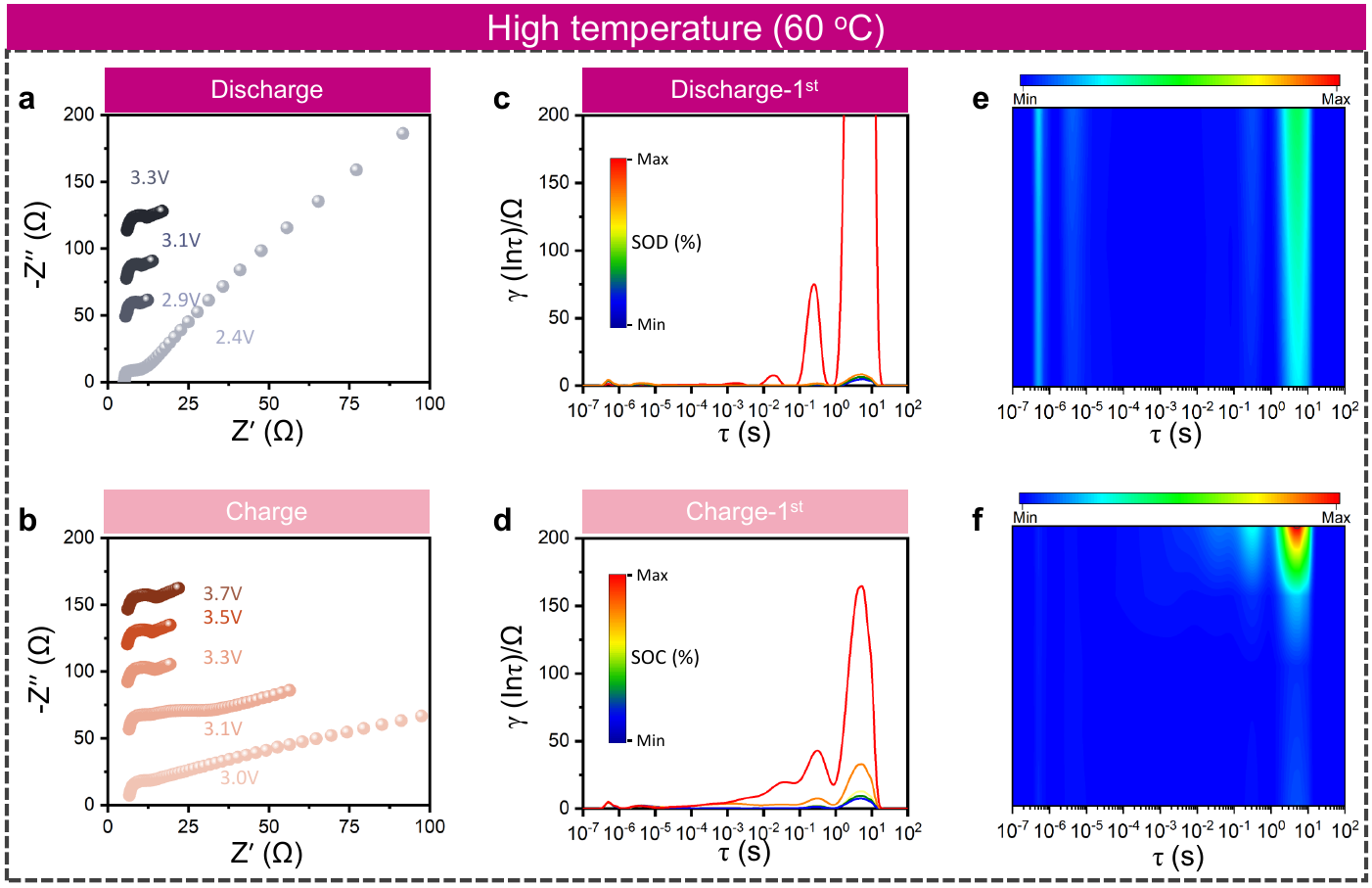}
\caption{High-temperature interfacial dynamics. 
Electrochemical characterization of LNO@NCM712/Li$_{6.6}$Si$_{0.6}$Sb$_{0.4}$S$_5$I/Li-In batteries at a high temperature of $60~^\circ$C during the initial cycle. Impedance spectra at various (a,b) charge (SOC) and discharge (SOD) states at high temperature, with corresponding (c,d) DRT curves and (e,f) 2D intensity maps, demonstrating thermal effects on interfacial kinetics and electrochemical behavior.
}
\label{Fig. S12}
\end{figure*}

\clearpage

\setcounter{table}{0}
\renewcommand{\thetable}{S\arabic{table}}

\begin{table*}[!ht]
\centering
\caption{Crystallographic data for Li$_6$SbS$_5$I. 
Rietveld analyses for the XRD pattern of Li$_6$SbS$_5$I. Space group $F$-$43m$, $a = 10.4104(1)$ (\AA), $R{\text{wp}} = 9.98\%$, confirming phase integrity and accurate structural modeling.
}
\label{tab:S1}
\setlength{\tabcolsep}{7.8mm}{}
\renewcommand{\arraystretch}{1.1}
\begin{tabular}{ccccc}
\hline
Atom site     & Wykoff site     & $x$           & $y$            & $z$              \\
\hline
Li1           & 48h             & 0.2845(1)     & 0.5232(1)      & 0.7845(1)        \\
I1            & 4g              & 0.5           & 0.5            & 0.0              \\
I2            & 4c              & 0.5           & 0.5            & 0.5              \\
Sb1           & 4b              & 0.5           & 0.5            & 0.5              \\
S1            & 4a              & 0.5           & 0.5            & 0.0              \\
S2            & 4c              & 0.25          & 0.25           & 0.25             \\
S3            & 16e             & 0.3719(1)     & 0.3719(1)      & 0.3719(1)        \\
\hline
\end{tabular}
\end{table*}

\clearpage

\begin{table*}[!ht]
\centering
\caption{Crystallographic data for Li$_{6.6}$Si$_{0.6}$Sb$_{0.4}$S$_5$I.
Rietveld analyses for the XRD pattern of Li$_{6.6}$Si$_{0.6}$Sb$_{0.4}$S$_5$I. Space group $F$-$43m$, $a = 10.3545(1)$ (\AA), $R{\text{wp}} = 9.13\%$, indicating high structural fidelity and phase purity.
}
\label{tab:S2}
\setlength{\tabcolsep}{7.8mm}{}
\renewcommand{\arraystretch}{1.1}
\begin{tabular}{ccccc}
\hline
Atom site     & Wykoff site     & $x$           & $y$            & $z$              \\
\hline
Li1           & 48h             & 0.2815(1)     & 0.5216(1)      & 0.7815(1)        \\
I1            & 4g              & 0.5           & 0.5            & 0.0              \\
Sb1           & 4b              & 0.5           & 0.5            & 0.5              \\
Si1           & 4c              & 0.5           & 0.5            & 0.5              \\
S1            & 4a              & 0.5           & 0.5            & 0.0              \\
S2            & 4c              & 0.25          & 0.25           & 0.25             \\
S3            & 16e             & 0.3754(1)     & 0.3754(1)      & 0.6246(1)        \\
\hline
\end{tabular}
\end{table*}

\clearpage

\begin{table*}[!ht]
\centering
\caption{Crystallographic data for Li$_{6.75}$Si$_{0.75}$Sb$_{0.25}$S$_5$I.
Rietveld analyses for the XRD pattern of Li$_{6.75}$Si$_{0.75}$Sb$_{0.25}$S$_5$I. Space group $F$-$43m$, $a = 10.3354(1)$ (\AA), $R{\text{wp}} = 10.95\%$, demonstrating precise lattice parameters and compositional tuning.}
\label{tab:S3}
\setlength{\tabcolsep}{7.8mm}{}
\renewcommand{\arraystretch}{1.1}
\begin{tabular}{ccccc}
\hline
Atom site     & Wykoff site      & $x$          & $y$            & $z$              \\
\hline
Li1           & 48h              & 0.2815(1)    & 0.5216(1)      & 0.7815(1)        \\
I1            & 4g               & 0.5          & 0.5            & 0.0              \\
Sb1           & 4b               & 0.5          & 0.5            & 0.5              \\
Si1           & 4c               & 0.5          & 0.5            & 0.5              \\
S1            & 4a               & 0.5          & 0.5            & 0.0              \\
S2            & 4c               & 0.25         & 0.25           & 0.25             \\
S3            & 16e              & 0.3754(1)    & 0.3754(1)      & 0.6246(1)        \\
\hline
\end{tabular}
\end{table*}

\clearpage

\begin{table*}[!ht]
\centering
\caption{Phase fractions of the main phase and impurity phases obtained from Rietveld refinement.}
\label{tab:S4}
\small
\setlength{\tabcolsep}{7.8mm}{}
\renewcommand{\arraystretch}{1.1}
\begin{tabular}{@{}lcccc@{}}
\hline
Sample & Main & Li$_2$S & LiI & Li$_4$SiS$_4$ \\
       & (wt\%) & (wt\%) & (wt\%) & (wt\%) \\
\hline
Li$_6$SbS$_5$I & 93.0 & 3.3 & 2.7 & 0.0 \\
Li$_{6.6}$Si$_{0.6}$Sb$_{0.4}$S$_5$I & 95.0 & 2.0 & 2.0 & 1.0 \\
Li$_{6.75}$Si$_{0.75}$Sb$_{0.25}$S$_5$I & 88.0 & 3.9 & 4.7 & 3.4 \\
\hline
\end{tabular}
\end{table*}

\clearpage

\begin{table*}[!ht]
\centering
\caption{Fitted EIS parameters and calculated ionic conductivities of Li$_6$SbS$_5$I.}
\label{tab:S5}
\begin{tabular}{lccccc}
\hline
Parameter & 30 $^\circ$C & 40 $^\circ$C & 50 $^\circ$C & 60 $^\circ$C & 70 $^\circ$C \\
\hline
$R_s$ / $\Omega$ & 94.980 & 85.154 & 72.603 & 127.820 & 9.516 \\
$R_{\mathrm{SE}}$ / $\Omega$ & 5068.027 & 4330.506 & 3090.115 & 2033.627 & 1933.228 \\
CPE$_{\mathrm{SE}}$ & 1.09E-10 & 1.16E-10 & 1.21E-10 & 1.26E-10 & 1.08E-10 \\
CPE$_{\mathrm{SS}}$ & 1.46E-07 & 1.30E-07 & 1.79E-07 & 2.30E-07 & 1.27E-07 \\
$\sigma$ / S cm$^{-1}$ & 1.50E-05 & 1.67E-05 & 1.96E-05 & 1.11E-05 & 1.50E-04 \\
$\chi^2$ & 3.47E-02 & 3.44E-02 & 3.07E-02 & 2.90E-02 & 3.23E-02 \\
\hline
\end{tabular}
\end{table*}
\clearpage

\begin{table*}[!ht]
\centering
\caption{Fitted EIS parameters and calculated ionic conductivities of Li$_{6.6}$Si$_{0.6}$Sb$_{0.4}$S$_5$I.}
\label{tab:S6}
\begin{tabular}{lccccc}
\hline
Parameter & 30 $^\circ$C & 40 $^\circ$C & 50 $^\circ$C & 60 $^\circ$C & 70 $^\circ$C \\
\hline
$R_s$ / $\Omega$ & 7.254 & 5.710 & 4.269 & 3.523 & 3.128 \\
$R_{\mathrm{SE}}$ / $\Omega$ & 8.123 & 5.895 & 5.206 & 4.198 & 3.112 \\
CPE$_{\mathrm{SE}}$ & 2.23E-05 & 2.69E-05 & 5.41E-05 & 1.69E-04 & 1.15E-05 \\
CPE$_{\mathrm{SS}}$ & 7.03E-06 & 8.81E-06 & 1.28E-05 & 1.84E-05 & 2.91E-05 \\
$\sigma$ / S cm$^{-1}$ & 9.90E-03 & 1.26E-02 & 1.68E-02 & 2.04E-02 & 2.30E-02 \\
$\chi^2$ & 5.14E-04 & 6.62E-04 & 1.88E-03 & 2.20E-03 & 3.04E-03 \\
\hline
\end{tabular}
\end{table*}

\clearpage

\begin{table*}[!ht]
\centering
\caption{Fitted EIS parameters and calculated ionic conductivities of Li$_{6.75}$Si$_{0.75}$Sb$_{0.25}$S$_5$I.}
\label{tab:S7}
\begin{tabular}{lccccc}
\hline
Parameter & 30 $^\circ$C & 40 $^\circ$C & 50 $^\circ$C & 60 $^\circ$C & 70 $^\circ$C \\
\hline
$R_s$ / $\Omega$ & 11.156 & 7.595 & 5.694 & 4.545 & 3.621 \\
$R_{\mathrm{SE}}$ / $\Omega$ & 15.338 & 7.607 & 7.139 & 5.762 & 3.264 \\
CPE$_{\mathrm{SE}}$ & 1.29E-05 & 6.27E-06 & 1.17E-05 & 1.17E-05 & 7.51E-06 \\
CPE$_{\mathrm{SS}}$ & 6.95E-06 & 9.21E-06 & 1.32E-05 & 2.10E-05 & 3.59E-05 \\
$\sigma$ / S cm$^{-1}$ & 8.10E-03 & 1.19E-02 & 1.59E-02 & 1.99E-02 & 2.50E-02 \\
$\chi^2$ & 6.39E-04 & 1.31E-03 & 1.78E-03 & 3.09E-03 & 5.44E-03 \\
\hline
\end{tabular}
\end{table*}

\clearpage

\begin{table*}[!ht]
\centering
\caption{Impedance parameters after cycling. 
Detailed impedance values for each component of LNO@NCM712/Li$_{6.6}$Si$_{0.6}$Sb$_{0.4}$S$_5$I/Li-In cells after cycling at 0.5C, providing quantitative insights into interfacial and bulk resistances following prolonged operation.
}
\label{tab:S8}
\setlength{\tabcolsep}{6.8mm}{}
\renewcommand{\arraystretch}{1.1}
\begin{tabular}{lclll}
\hline
Element                         & Parameter                & \multicolumn{3}{c}{Value (After cycle) long-cycling performance}        \\
\cline{3-5}
                                &                          & $x = 0$                 & $x = 0.2$                & $x = 0.3$          \\
\hline
$R_{\text{SE\_bulk}}$           & $R$                      & 39.4 $\Omega$           & 15.5 $\Omega$            & 30.1 $\Omega$      \\
$R_{\text{cathode/electro}}$    & $R$                      & 16.5 $\Omega$           & 8.8 $\Omega$             & 19.3 $\Omega$      \\
$C_{\text{eq}}$                 &                          & 1.88 mF                 & 1.3 mF                   & 0.68 mF            \\
$\alpha$                        &                          & 216 m                   & 358 m                    & 481 m              \\
$R$                             &                          & 357 $\Omega$            & 121 $\Omega$             & 247 $\Omega$       \\
$C_{\text{anode/electro}}$      & $R$                      & 303 $\mu$F              & 632 nF                   & 1.44 mF            \\
$\alpha$                        &                          & 915 m                   & 459 m                    & 829 m              \\
$R_{\text{diffusion}}$          & $W$                      & 841 DW                  & 16.4 DW                  & 7.59 DW            \\
\hline
\end{tabular}
\end{table*}

\clearpage

\begin{table*}[!ht]
\centering
\caption{Bader charge and corresponding charge difference for cathodes with varying carbon amounts (x = 0.0, 0.2, 0.3). TBC denotes the total Bader charge, while CD represents the charge difference.}
\label{tab:S9}
\setlength{\tabcolsep}{6.2mm}{}
\renewcommand{\arraystretch}{1.1}
\begin{tabular}{cllllll}
\hline
\hline
          &      \multicolumn{2}{c}{x = 0.0}     &      \multicolumn{2}{c}{x = 0.2}           &      \multicolumn{2}{c}{x = 0.3}              \\
Atom      & TBC              & CD                & TBC                       & CD             & TBC                       & CD                \\
\hline
Li        & 33.00            & —                 & 23.28                     & -9.72          & 23.35                     & -9.65             \\
Ni        & 100.00           & —                 & 88.56                     & -11.44         & 89.02                     & -10.98            \\
Mn        & 26.00            & —                 & 22.31                     & -3.69          & 22.38                     & -3.62             \\
O         & 144.00           & —                 & 169.87                    & 25.87          & 169.36                    & 25.36             \\
Co        & 9.00             & —                 & 7.73                      & -1.27          & 7.46                      & -1.54             \\
C         & —                & —                 & 64.27                     & 0.27           & 96.43                     & 0.43              \\
\hline
\hline
\end{tabular}
\end{table*}

\clearpage

\end{document}